\documentclass[twocolumn, 10pt, pre, aps, showpacs, reprint, amsmath, amssymb, superscriptaddress, nofootinbib]{revtex4-1}

\usepackage[]{graphicx}
\usepackage{subfigure}
\usepackage{units}

\begin{document}
 
\newcommand{\kt}{k_\parallel}
\newcommand{\lzt}{q_z}
\newcommand{\atled}{\bm{\nabla}}
\newcommand{\dx}{\frac{\partial}{\partial_x}}
\newcommand{\dy}{\frac{\partial}{\partial_y}}
\newcommand{\dz}{\frac{\partial}{\partial_z}}
\newcommand{\dt}{\frac{\partial}{\partial_t}}
\newcommand{\sqrdt}{\frac{\partial^2}{\partial_t^2}}
\newcommand{\pbyp}[2]{\frac{\partial #1}{\partial #2}}
\newcommand{\dbyd}[2]{\frac{d #1}{d #2}}
\newcommand{\ex}{\vec{e}_x}
\newcommand{\ey}{\vec{e}_y}
\newcommand{\ez}{\vec{e}_z}
\newcommand{\besselj}[2]{\mathrm{J}_{#1}(#2)}
\newcommand{\besseljp}[2]{\mathrm{J'}_{#1}(#2)}
\newcommand{\besseljs}[1]{\mathrm{J}_{#1}}
\newcommand{\besseljsp}[1]{\mathrm{J}_{#1}'}
\newcommand{\besseljspp}[1]{\mathrm{J}_{#1}''}
\newcommand{\hankel}[3]{\mathrm{H}_{#1}^{(#2)}(#3)}
\newcommand{\hankelp}[3]{\mathrm{H'}_{#1}^{(#2)}(#3)}
\newcommand{\hankels}[2]{\mathrm{H}_{#1}^{(#2)}}
\newcommand{\hankelsp}[2]{\mathrm{H}_{#1}'^{(#2)}}
\newcommand{\hankelspp}[2]{\mathrm{H}_{#1}''^{(#2)}}
\newcommand{\laplace}{\Delta}
\newcommand{\neff}{n_{\mathrm{eff}}}
\newcommand{\fexp}{f_{\mathrm{expt}}}
\newcommand{\ftheo}{f_{\mathrm{calc}}}
\newcommand{\nexp}{\tilde{n}}
\newcommand{\Gtheo}{\Gamma_{\mathrm{calc}}}
\newcommand{\Gexp}{\Gamma_{\mathrm{expt}}}
\newcommand{\Grad}{\Gamma_{\mathrm{rad}}}
\newcommand{\Gabs}{\Gamma_{\mathrm{abs}}}
\newcommand{\Gant}{\Gamma_{\mathrm{ant}}}
\newcommand{\rhof}{\rho_{\mathrm{fluc}}}
\newcommand{\rhofscl}{\rhof^{\mathrm{scl}}}
\newcommand{\rhofpo}{\rho_{\ppo, r}}
\newcommand{\rhow}{\rho_{\mathrm{Weyl}}}
\newcommand{\Nweyl}{N_{\mathrm{Weyl}}}
\newcommand{\rhot}{\tilde{\rho}}
\newcommand{\rhotscl}{\rhot_{\mathrm{scl}}}
\newcommand{\rhotpo}{\rhot_{\ppo, r}}
\newcommand{\fmin}{f_{\mathrm{min}}}
\newcommand{\kmin}{k_{\mathrm{min}}}
\newcommand{\fmax}{f_{\mathrm{max}}}
\newcommand{\kmax}{k_{\mathrm{max}}}
\newcommand{\Dk}{\Delta k}
\newcommand{\fcrit}{f_{\mathrm{crit}}}
\newcommand{\kcrit}{k_{\mathrm{crit}}}
\newcommand{\po}{\mathrm{po}}
\newcommand{\ppo}{p}
\newcommand{\lpo}{\ell_\po}
\newcommand{\lppo}{\ell_\ppo}
\newcommand{\lpeak}{\ell_\mathrm{peak}}
\newcommand{\lmax}{\ell_{\mathrm{max}}}
\newcommand{\alphacrit}{\alpha_{\mathrm{crit}}}
\newcommand{\chico}{\chi_{\mathrm{co}}}
\newcommand{\reffig}[1]{\mbox{Fig.~\ref{#1}}}
\newcommand{\subreffig}[1]{\mbox{Fig. \subref{#1}}}
\newcommand{\refeq}[1]{\mbox{Eq.~(\ref{#1})}}
\newcommand{\refsec}[1]{\mbox{Sec.~\ref{#1}}}
\newcommand{\reftab}[1]{\mbox{Table \ref{#1}}}
\newcommand{\etal}{\textit{et al.\ }}
\newcommand{\FSR}{f_\mathrm{FSR}}
\newcommand{\FSRmean}{\left< \FSR \right>}
\newcommand{\dist}{D}
\newcommand{\lchar}{l}
\newcommand{\Mon}{\mathbf{M}}
\renewcommand{\Re}[1]{\mathrm{Re}\left(#1\right)}
\renewcommand{\Im}[1]{\mathrm{Im}\left(#1\right)}
\hyphenation{re-so-nan-ce re-so-nan-ces ex-ci-ta-tion z-ex-ci-ta-tion di-elec-tric ap-pro-xi-ma-tion ra-dia-tion Me-cha-nics quan-tum pro-posed Con-cepts pro-duct Reh-feld ob-ser-va-ble Se-ve-ral rea-so-nable Ap-pa-rent-ly re-pe-ti-tions re-la-tive quan-tum su-per-con-duc-ting ap-pro-xi-mate cri-ti-cal}

\title{Trace formula for chaotic dielectric resonators tested with microwave experiments}

\author{S. Bittner}
\author{B. Dietz}
\email{dietz@ikp.tu-darmstadt.de}
\affiliation{Institut f\"ur Kernphysik, Technische Universit\"at Darmstadt, D-64289 Darmstadt, Germany}
\author{R. Dubertrand}
\affiliation{School of Mathematics, University of Bristol, University Walk, Bristol BS8 1TW, United Kingdom}
\author{J. Isensee}
\author{M. Miski-Oglu}
\affiliation{Institut f\"ur Kernphysik, Technische Universit\"at Darmstadt, D-64289 Darmstadt, Germany}
\author{A. Richter}
\email{richter@ikp.tu-darmstadt.de}
\affiliation{Institut f\"ur Kernphysik, Technische Universit\"at Darmstadt, D-64289 Darmstadt, Germany}
\affiliation{ECT*, Villa Tambosi, I-38123 Villazano (Trento), Italy}

\date{\today}

\begin{abstract} 
We measured the resonance spectra of two stadium-shaped dielectric microwave resonators and tested a semiclassical trace formula for chaotic dielectric resonators proposed by Bogomolny \textit{et al.} [Phys. Rev. E \textbf{78}, 056202 (2008)]. We found good qualitative agreement between the experimental data and the predictions of the trace formula. Deviations could be attributed to missing resonances in the measured spectra in accordance with previous experiments [Phys. Rev. E \textbf{81}, 066215 (2010)]. The investigation of the numerical length spectrum showed good qualitative and reasonable quantitative agreement with the trace formula. It demonstrated, however, the need for higher-order corrections of the trace formula. The application of a curvature correction to the Fresnel reflection coefficients entering the trace formula yielded better agreement, but deviations remained, indicating the necessity of further investigations.
\end{abstract}


\maketitle

\section{\label{sec:intro}Introduction}
The development of dielectric microresonators with their manifold applications \cite{Nockel2003, Vahala2004, Matsko2005}, such as microlasers \cite{McCall1992}, has triggered much interest in their theoretical description. Dielectric resonators are governed by the vectorial Helmholtz equation, whose exact treatment is generally not achievable. However, semiclassical methods have been applied successfully to the modeling of, e.g., the far-field patterns of microlasers \cite{Noeckel1994, Noeckel1997, Altmann2009} or the localization of their resonant modes, so-called scarring \cite{Rex2002, Gmachl2002, Harayama2003, Fang2005a, Fang2007, Fang2007a}. These involve the periodic orbits (POs) of the corresponding classical system, a two-dimensional (2D) dielectric billiard in the case considered here. The POs are, for example, connected to the density of states via trace formulas \cite{Gutzwiller1970, Gutzwiller1971, Brack2003}, and thus, also to the spectral properties of the corresponding resonator. Trace formulas for 2D dielectric resonators have been proposed \cite{Bogomolny2008, Hales2011} and tested experimentally with microlasers \cite{Lebental2007, Bogomolny2011} and microwave resonators \cite{Bittner2010, Bittner2011a, Bittner2012a} for various geometries, which predominantly correspond to dielectric billiards with regular classical dynamics. The aim of the work presented here has been a thorough experimental test of the trace formula for chaotic 2D dielectric resonators. We used two microwave resonators in the shape of stadia with different aspect ratios. The stadium billiard is a fully chaotic system \cite{Bunimovich1979} and has been investigated thoroughly theoretically \cite{McDonald1979, Christoffel1986, Bogomolny1988, Shudo1990, Meiss1992, Sieber1993, Alonso1994, Primack1994, Tanner1997, Casati1999} and experimentally \cite{Stoeckmann1990, Graf1992, Marcus1992, Alt1995, Stein1995, Crommie1995, Arcos1998, Friedman2001} in the context of quantum chaos. Stadium-shaped microlasers have been studied experimentally in Refs.~\cite{Fukushima2004, Fang2005a, Lebental2006, Fang2007, Fang2007a, Shinohara2008, Djellali2009}, and in Ref.~\cite{Bogomolny2011}, the length spectrum of a stadium-shaped microlaser was determined from measurements. However, there still was lack of a detailed comparison of the experimental length spectrum of a chaotic dielectric cavity with the trace formula prediction of Ref.~\cite{Bogomolny2008}. This has been the motivation of the present work. \newline
The article is organized as follows. Section~\ref{sec:expSetup} describes the experimental setup and the measured frequency spectrum, \refsec{sec:traceForm} summarizes the salient features of the trace formula for chaotic dielectric resonators, and \refsec{sec:expLspekt} contains the comparison of the experimental length spectra with the trace formula predictions. In \refsec{sec:numLspekt}, the measured data and the trace formula predictions are compared to numerically calculated data, and a correction to the trace formula at curved boundaries is investigated. Section~\ref{sec:concl} closes with some concluding remarks.

\section{\label{sec:expSetup}Experimental setup and frequency spectrum}

\begin{figure}[tb]
\begin{center}
\subfigure[]
{
	\includegraphics[width = 8.4cm]{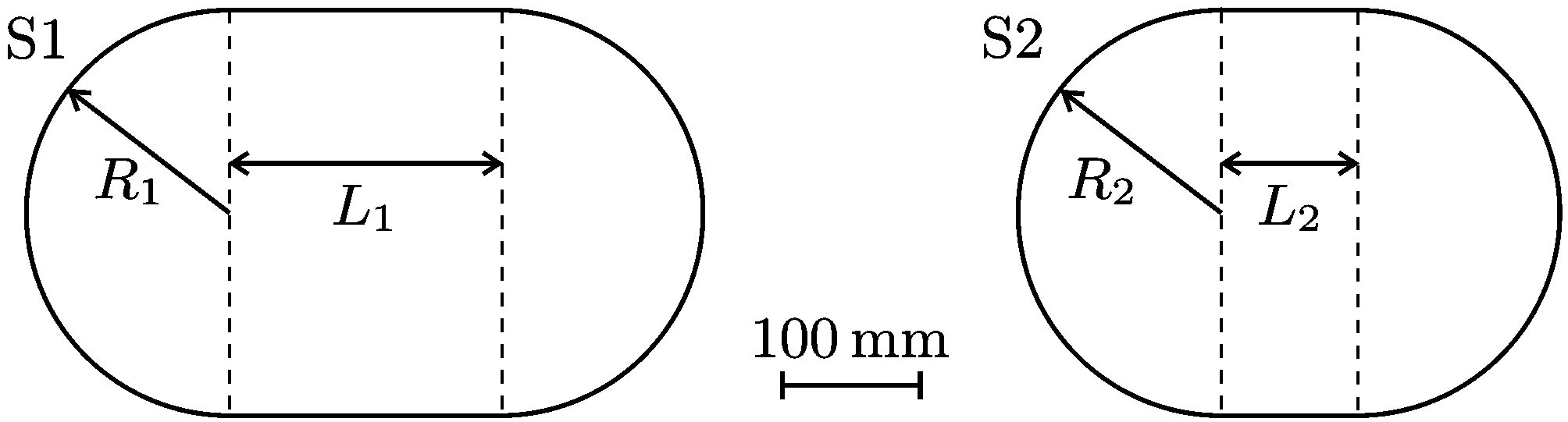}
	\label{sfig:setupTop} 
}
\subfigure[]
{ 
	\includegraphics[width = 8.4cm]{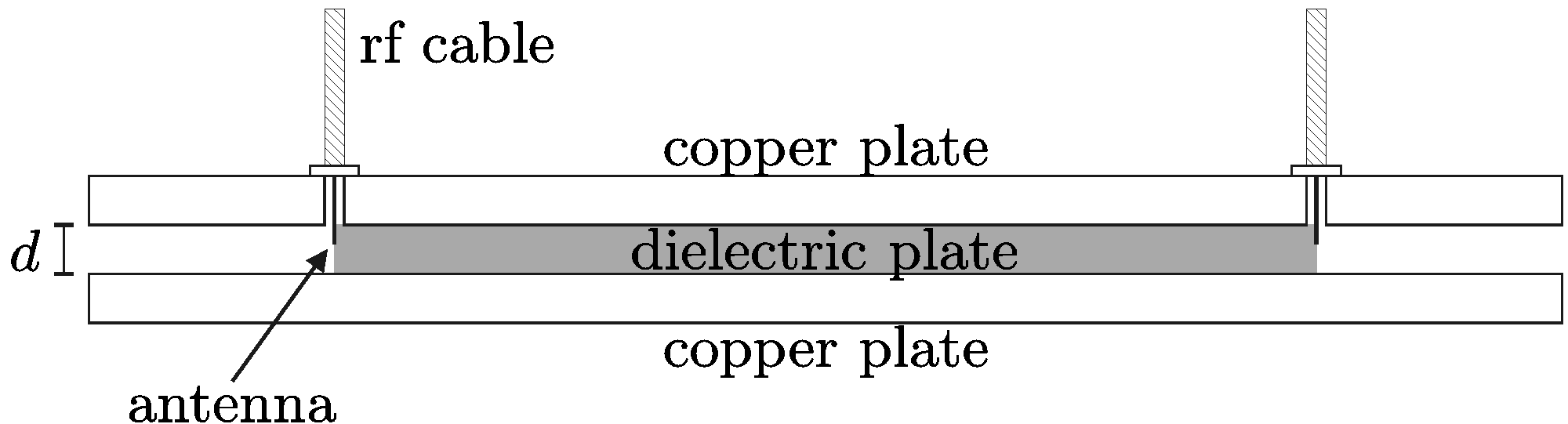}
	\label{sfig:setupSide}
}
\end{center}
\caption{\label{fig:setup}\subref{sfig:setupTop} Geometry of the two Teflon plates S1 (left) and S2 (right). The solid lines indicate the boundaries of the plates and the dashed lines the common border of the semicircular and the rectangular parts of the stadia, where $R$ is the radius of the semicircles and $L$ the length of the rectangular part. \subref{sfig:setupSide} Schematic side view (section) of the experimental setup (not to scale). The Teflon plate with thickness $d$ is put between two copper plates. Two antennas are led through small holes in the top copper plate next to the sidewalls of the Teflon plate. Reprinted from Ref.~\cite{Bittner2010}.}
\end{figure}

\begin{figure*}[tb]
\begin{center}
\includegraphics[width = 14 cm]{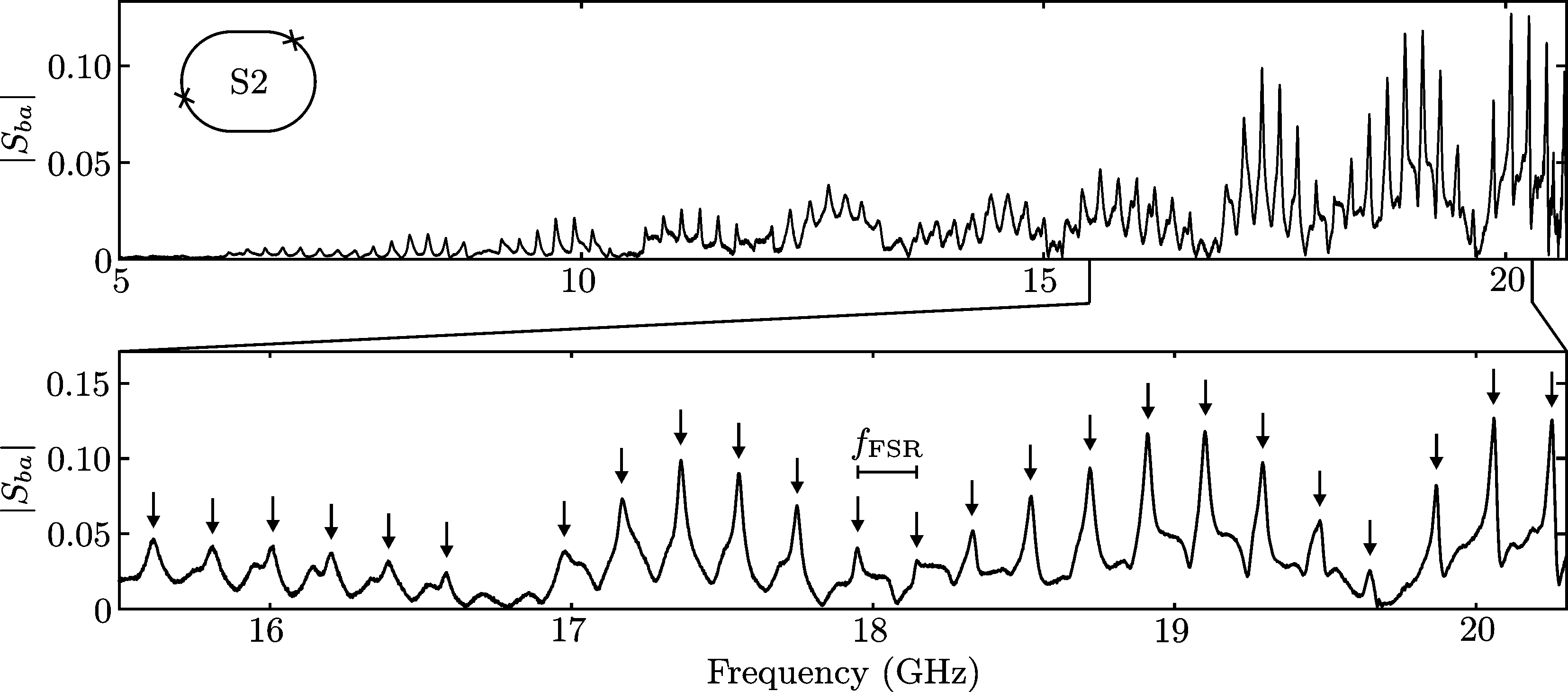}
\end{center}
\caption{The measured frequency spectrum of the stadium S2. The inset in the top part shows the shape of the stadium with the positions of the antennas indicated by crosses. The arrows in the bottom panel indicate a subset of resonances that are approximately equidistant, i.e., they have a constant free spectral range $\FSR$.}
\label{fig:freqSpec}
\end{figure*}

\begin{figure}[tb] 
\begin{center}
\includegraphics[width = 8.4 cm]{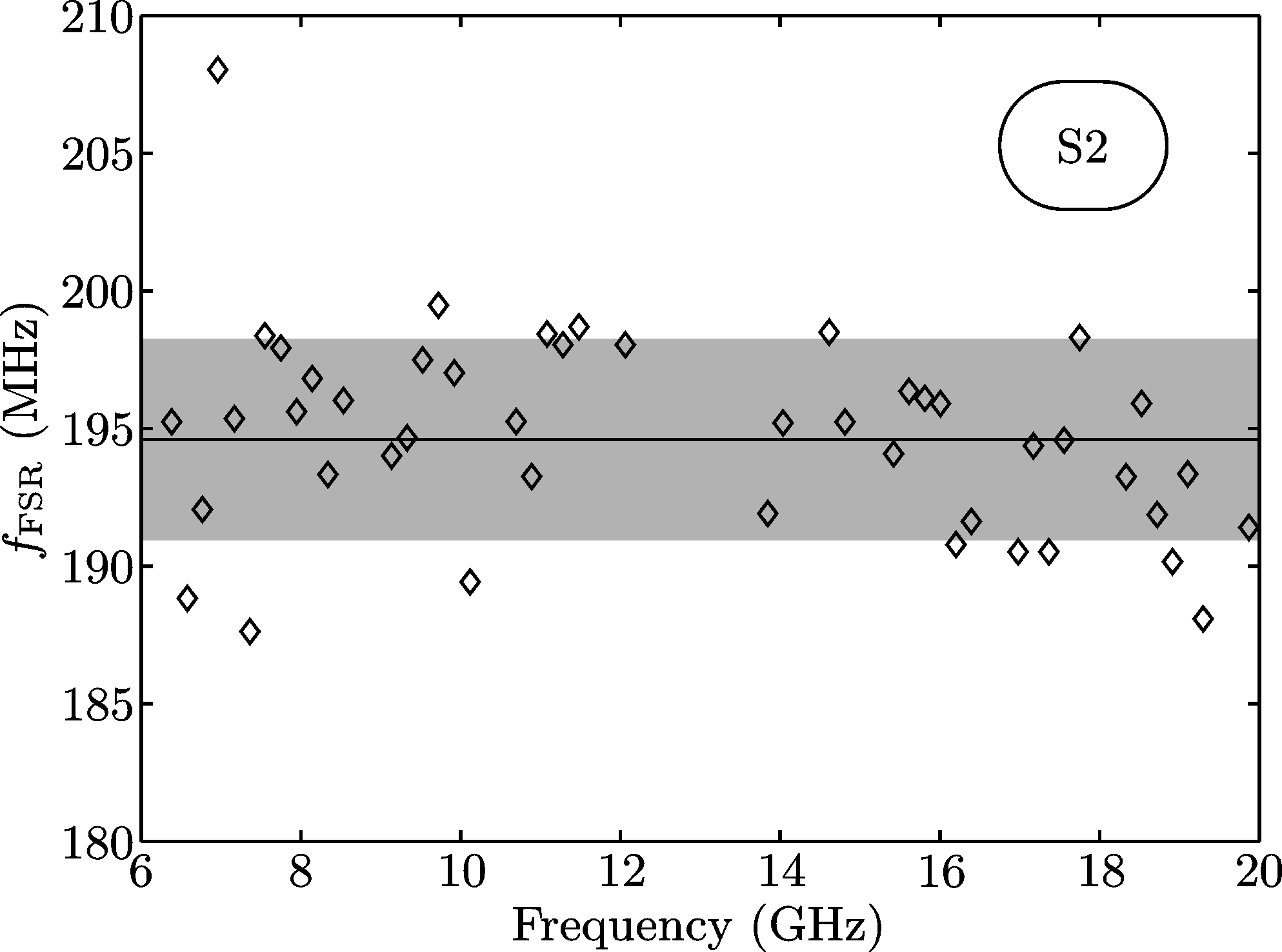}
\end{center}
\caption{Measured free spectral range (FSR) of the subset of equidistant resonances (marked by arrows in the bottom panel of \reffig{fig:freqSpec}) of resonator S2 with respect to the frequency. The horizontal line is the mean value of the measured FSR, and the gray bar indicates the standard deviation.}
\label{fig:equid} 
\end{figure}
  
Two flat Teflon plates in the shape of a Bunimovich stadium were used as microwave resonators. The geometry of the plates is shown in \reffig{sfig:setupTop}. The radius of the semicircle of the first stadium, designated S1, was $R_1 = \unit[150.1]{mm}$ and the length of its straight edge was $L_1 = \unit[200.2]{mm}$. Its  aspect ratio was thus $\epsilon_1 = L_1 / (2 R_1) = 0.67$. It had an index of refraction of $n_1 = 1.425$. The second stadium S2 had the parameters $R_2 = \unit[149.9]{mm}$, $L_2 = \unit[99.8]{mm}$, $\epsilon = 0.33$, and $n_2 = 1.404$. The index of refraction was determined in both cases with the same technique as in Ref.~\cite{Bittner2010}. The small difference between $R_1$ and $R_2$ is due to manufacturing uncertainties. The frequency range of interest, $5$--$20$ GHz, corresponds to $kR = 15.7$--$62.9$, where $k$ is the wave number. Both Teflon plates had a thickness of $d = \unit[5]{mm}$. The dielectric plates were put between two copper plates as shown in \reffig{sfig:setupSide}. Below the frequency
\begin{equation} f_{2D} = \frac{c}{2 n d} \, , \end{equation} 
only two-dimensional transverse magnetic modes exist in the resonator, where $c$ is the speed of light in vacuum and $n$ is the index of refraction \cite{Jackson1999, Sridhar1992}. In the cases considered here, $f_{2D} \approx \unit[20]{GHz}$. The setup is similar to that used in Ref.~\cite{Bittner2010}. Two vertical wire antennas were used to couple microwave power into and out of the resonator. They were placed next to the sidewalls of the Teflon plates as shown in \reffig{sfig:setupSide} and protruded $\unit[1]{mm}$ into the space between the copper plates. A vectorial network analyzer (PNA 5230A by Agilent Technologies) was connected to the antennas via coaxial rf cables. It measured the complex transmission amplitude $S_{ba}(f)$ from antenna $a$ to antenna $b$ for a given frequency $f$. Its modulus squared is given by
\begin{equation} |S_{ba}|^2 = \frac{P_{b, \mathrm{out}}}{P_{a, \mathrm{in}}} \, , \end{equation}
where $P_{a, \mathrm{in}}$ is the power coupled into the resonator via antenna $a$ and $P_{b, \mathrm{out}}$ that coupled out via antenna $b$. The plot of $|S_{ba}(f)|^2$ versus the frequency yields the transmission spectrum of the resonator. \newline
Figure \ref{fig:freqSpec} shows the measured frequency spectrum of stadium S2. It comprises a multitude of resonances of diverse widths on top of a slowly oscillating background that is attributed to direct transmission processes between the antennas. The quality factors $Q_j = f_j / \Gamma_j$ of the resonances, where $f_j$ is the resonance frequency and $\Gamma_j$ the full width at half maximum of resonance $j$, are in the range of $100$--$1200$, with a mean value of $\left< Q \right> \approx 300$. The relatively low quality factors are attributed to the large radiation losses of the stadia. The resonances are additionally broadened due to absorption in the Teflon, Ohmic losses in the copper plates, and the power coupled out by the antennas. The resonance frequencies and widths were determined by fitting Lorentzians to the measured spectrum. The frequency spectrum includes a series of apparently equidistant resonances that are indicated by the arrows in the bottom panel of \reffig{fig:freqSpec}. Similar structures were observed in the lasing spectra of stadium-shaped polymer ($n = 1.5$) microlasers \cite{Lebental2006, Djellali2009, Bogomolny2011}. The free spectral range (FSR) of this series, i.e., the frequency spacing $\FSR$ between adjacent modes, is plotted in \reffig{fig:equid}. The FSR is indeed constant over the whole frequency range considered here, with a mean value of $\FSRmean = 194.6$ MHz within a standard deviation of $\Delta \FSR = 3.7$ MHz. The fluctuations of the measured $\FSR$ can at least partly be attributed to the error in the determination of the resonance frequencies when modes strongly overlap, as is generally the case for the spectra presented in this work (cf.~\reffig{fig:freqSpec}). Other, systematic reasons for the fluctuations of the FSR are not known but cannot be excluded. Also, for stadium S1 such a series of equidistant resonances was found with $\FSRmean = (165.4 \pm 3.8)$ MHz. This leads to the presumption that the corresponding resonant modes are localized around one or several POs (so-called scarring \cite{Heller1984}), as has been observed for various dielectric cavities \cite{Rex2002, Gmachl2002, Harayama2003, Fang2005a, Fang2007, Fang2007a}. The length of the underlying PO(s) can be determined from the FSR via \cite{Lebental2007}
\begin{equation} \label{eq:lchar} \lchar = \frac{c}{n \, \FSRmean} \, . \end{equation}
This approach is commonly used to identify the POs that play the dominant role for the modes of a microlaser, though a clear identification is not always possible \cite{Gmachl2002, Chern2004, Kim2011, Kim2011a, Shinohara2011b, Yi2011}. From \refeq{eq:lchar} we obtain $\lchar_1 = (1.272 \pm 0.029)$ m for stadium S1 and $\lchar_2 = (1.097 \pm 0.021)$ m for stadium S2. We will call $\lchar$ the path length in the following and compare it to the lengths of the various POs in \refsec{sec:expLspekt}. The path length $\lchar_1$ corresponds to $94.8\%$ of the circumference of stadium S1 and $\lchar_2$ to $96.1\%$ of that of stadium S2, indicating that the modes belonging to the equidistant series are of the whispering gallery type. This supposition is supported by the fact that the associated resonances are the ones with the highest quality factors.

\section{\label{sec:traceForm}Helmholtz equation and trace formula}
Flat microlasers or dielectric microwave resonators can only be treated approximately as 2D systems by introducing an effective index of refraction. This approximation, however, has only a limited accuracy, as discussed in Refs.~\cite{Bittner2009, Bittner2012a}. Therefore the Teflon plates were squeezed between two metallic plates as depicted in \reffig{sfig:setupSide}. Such resonators are described by the 2D scalar Helmholtz equation \cite{Jackson1999}
\begin{equation} \label{eq:helmholtz} \left[ \Delta + n^2(\vec{r}) k^2 \right] E_z = 0 \end{equation}
for frequencies $f \leq f_{2D}$. Here, $n(\vec{r})$ is the index of refraction at position $\vec{r}$, $k = 2 \pi f / c$ is the wave number, and $E_z$ is the $z$ component of the electric field strength. Outgoing wave boundary conditions are imposed on $E_z$ to account for the openness of the resonator \cite{Nockel2003}. Therefore, the eigenvalues $k_j$ of \refeq{eq:helmholtz} are complex, where $f_j = c \, \Re{k_j} / (2 \pi)$ is the frequency and $\Gamma_j = -c \, \Im{k_j} / \pi$ is the width of the resonance $j$ (see \reffig{fig:freqSpec}). The width $\Gamma_j$ accounts for the radiation losses of the cavity. In a microwave experiment, the power coupled out of the resonator by the antennas and that absorbed by the dielectric material and the metal plates results in an increase of $\Gamma_j$.  \newline 
Trace formulas relate the density of states (DOS) of a wave-dynamical system to the POs of the corresponding classical system \cite{Gutzwiller1970, Gutzwiller1971, Brack2003}. For an open dielectric resonator the DOS is
\begin{equation} \label{eq:dos} \rho(k) = - \frac{1}{\pi} \sum \limits_j \frac{\Im{k_j}}{[k - \Re{k_j}]^2 + [\Im{k_j}]^2} \, . \end{equation}
It can be decomposed into a smooth, average part $\bar{\rho}(k)$, and a fluctuating part $\rhof(k)$, i.e., $\rho = \bar{\rho} + \rhof$. The smooth part is related to the area $A$ and the circumference $U$ of the resonator via the Weyl law given in Ref.~\cite{Bogomolny2008}. In the semiclassical limit, the fluctuating part $\rhof$ can be written as a sum over contributions from the POs of the corresponding dielectric billiard. For a fully chaotic billiard like the stadium, i.e., a billiard for which all POs are unstable and isolated, the trace formula proposed in Ref.~\cite{Bogomolny2008} is based on the well-known Gutzwiller trace formula \cite{Gutzwiller1971, Brack2003}. It is 
\begin{equation} \label{eq:trFormChaot} \rhofscl(k) = \sum_\ppo \sum \limits_{r = 1}^\infty \rhofpo(k) + \textrm{c.c.} \, , \end{equation}
with the contribution of the $r$th repetition of the primitive periodic orbit $\ppo$ being
\begin{equation} \label{eq:trFormPrim} \rhofpo(k) = \frac{n \lppo}{\pi |\det(\Mon_\ppo^r - 1)|^{1/2}} \, R_\ppo^r \, e^{i\{r n k \lppo - r \mu_\ppo \pi/2\}} \, . \end{equation}
Here, $\lppo$ is the length of the primitive periodic orbit, $\Mon_\ppo$ is the monodromy matrix characterizing its stability, $R_\ppo = \prod_j r_j$ is the product of the Fresnel reflection coefficients $r_j$ for each reflection of the primitive PO at the boundary, and $\mu_\ppo$ is the Maslov index, counting the number of focal points and caustics. The essential difference to the Gutzwiller trace formula is the additional factor $R_\ppo$ accounting for the dielectric boundary conditions. Since the polarization of the electric field, $\vec{E} = E_z \ez$, is perpendicular to the plane of incidence of the POs, the Fresnel reflection coefficients are given by \cite{Hecht2002}
\begin{equation} \label{eq:fresnelPlan} r_j = \frac{n \cos{\theta_j} - \sqrt{1 - n^2 \sin^2{\theta_j}}}{n \cos{\theta_j} + \sqrt{1 - n^2 \sin^2{\theta_j}}}  \, , \end{equation}
where $\theta_j$ is the angle of incidence with respect to the boundary normal for the $j$th reflection. The POs of the billiards were obtained with the algorithm of Ref.~\cite{Biham1992}, their monodromy matrices $\Mon_\ppo$ were calculated following Ref.~\cite{Sieber1990a}, and the Maslov indices $\mu_\ppo$ using Ref.~\cite{Lin1997}. The trace formula [\refeq{eq:trFormChaot}] is only valid for isolated, unstable POs. In the stadium billiard, a family of marginally stable POs which bounce between the parallel parts of the billiard (bouncing ball orbits, BBOs) exists that needs a separate treatment \cite{Graf1992, Sieber1993, Primack1994, Tanner1997, Alonso1994}. In our case, however, these orbits are of little relevance since the reflection coefficients for vertical incidence are quite small. \newline
Instead of studying the DOS itself, the Fourier transform (FT) of its fluctuating part was considered,
\begin{equation} \label{eq:FTdef} \rhot(\ell) = \int_{\kmin}^{\kmax} dk \, \rhof(k) e^{-i k n \ell} \, = \mathcal{F} \{ \rhof \}, \end{equation}
where $[\kmin, \kmax] = 2 \pi [\fmin, \fmax] / c$ is the wave num\-ber, i.e., the fre\-que\-ncy interval under consideration, and $\ell$ is a geometrical length, and compared to the FT of the semiclassical trace formula, \refeq{eq:trFormChaot}. From \refeq{eq:dos} follows
\begin{equation} \label{eq:lspektExp} \rhot(\ell) = \sum \limits_j e^{- i k_j n \ell} - \mathcal{F}\{ \bar{\rho} \} \, , \end{equation}
and $|\rhot(\ell)| = |\mathcal{F}\{\rhof(k)\}|$ is called the length spectrum. The experimental length spectrum $|\rhot(\ell)|$ is computed by inserting the resonance frequencies $f_j$ and the widths $\Gamma_j$ obtained from the measured frequency spectrum into \refeq{eq:lspektExp}. The FT of the contribution of a single PO to the trace formula [\refeq{eq:trFormPrim}] is
\begin{equation} \label{eq:singlePOcontr} \rhotpo(\ell) = a_{\ppo, r} \, \mathrm{sinc}[n \Dk (r \lppo - \ell) /2 ] \end{equation}
with $\mathrm{sinc}(x) = \sin{(x)} / x$ and
\begin{equation} a_{\ppo, r} = \frac{\Dk \, n \lppo}{\pi |\det{(\Mon_\ppo^r - 1)}|^{1/2}} R_\ppo^r e^{i \{ n \bar{k} (r \lppo - \ell) - r \mu_\ppo \pi / 2 \}} \, , \end{equation}
where $\Dk = \kmax - \kmin$ and $\bar{k} = (\kmax + \kmin) / 2$. Furthermore, each PO is counted several times in \refeq{eq:trFormChaot} depending on its symmetry with respect to the two mirror symmetry axes of the stadium and with respect to time reversal. For each symmetry that is lacking for a certain PO, its contribution is counted twice \cite{Robbins1989, Takami1995, Bruus1996}, which leads to a symmetry factor $s_{\ppo} \in \{1, 2, 4, 8\}$. It should be noted that the FT of the DOS is different from that of the transmission amplitude $S_{ba}$: while the former (i.e., the length spectrum) is connected to the POs of the billiard, the latter is proportional to the propagator from antenna $a$ to antenna $b$ and is related to the classical trajectories between the antennas \cite{Stein1995, Berry1972}.

\section{\label{sec:expLspekt}The experimental length spectra}

\begin{figure*}[tb]
\begin{center}
\includegraphics[width = 12 cm]{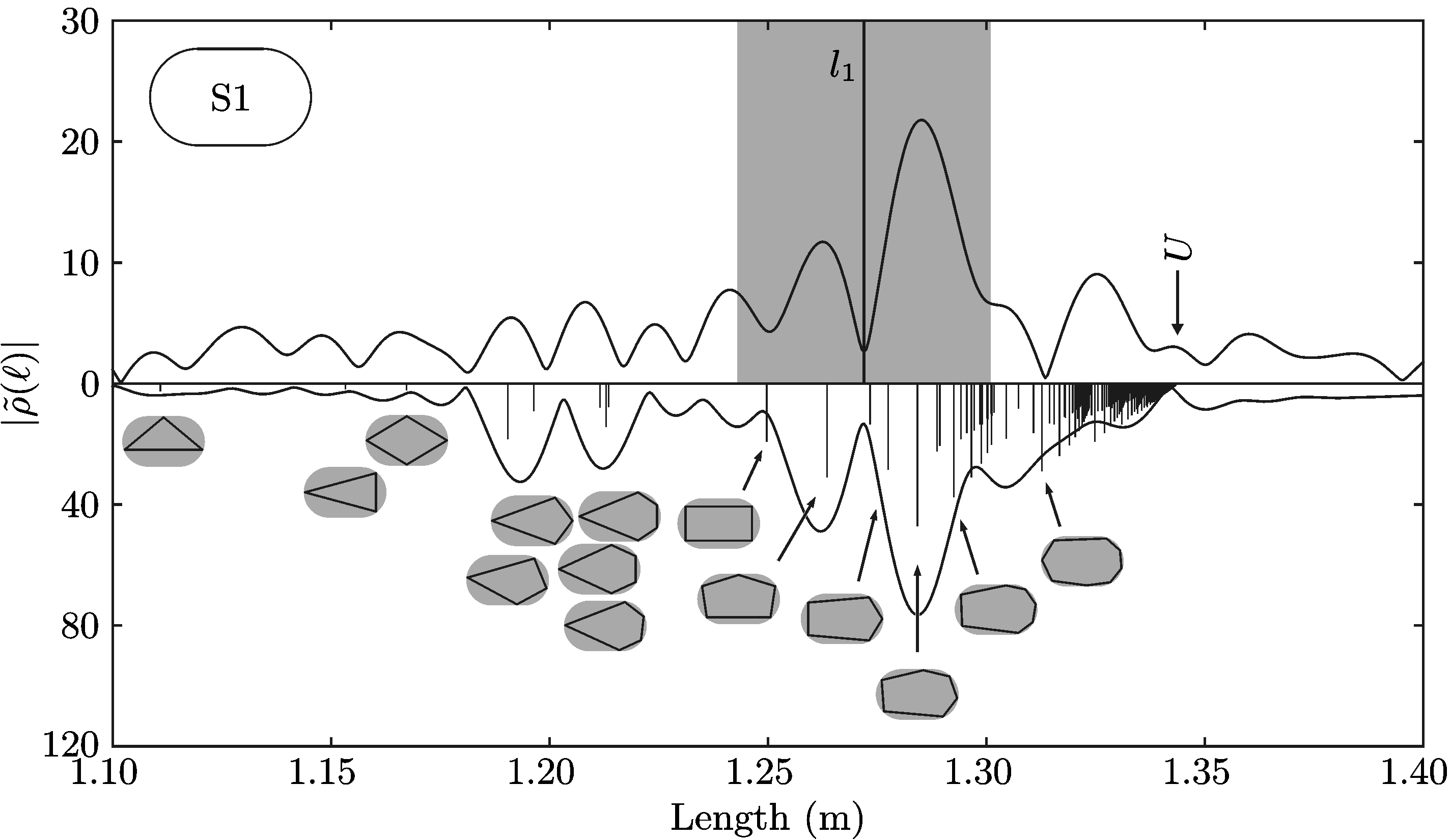}
\end{center}
\caption{\label{fig:lspectS1}Length spectrum for the stadium S1. The top graph shows the experimental length spectrum and the bottom graph the FT of the trace formula. Note the different scales of the top and bottom graphs. The vertical line in the top graph indicates the path length $\lchar_1$ defined in \refeq{eq:lchar}, the gray bar its error, and the arrow the circumference $U$ of the stadium. The vertical lines in the bottom graph indicate the lengths and the amplitudes $A_{\ppo, r}$ of the POs [\refeq{eq:poAmp}] used for the calculation of the trace formula. Some POs are shown as insets.}
\end{figure*}

\begin{figure*}[tb]
\begin{center}
\includegraphics[width = 12 cm]{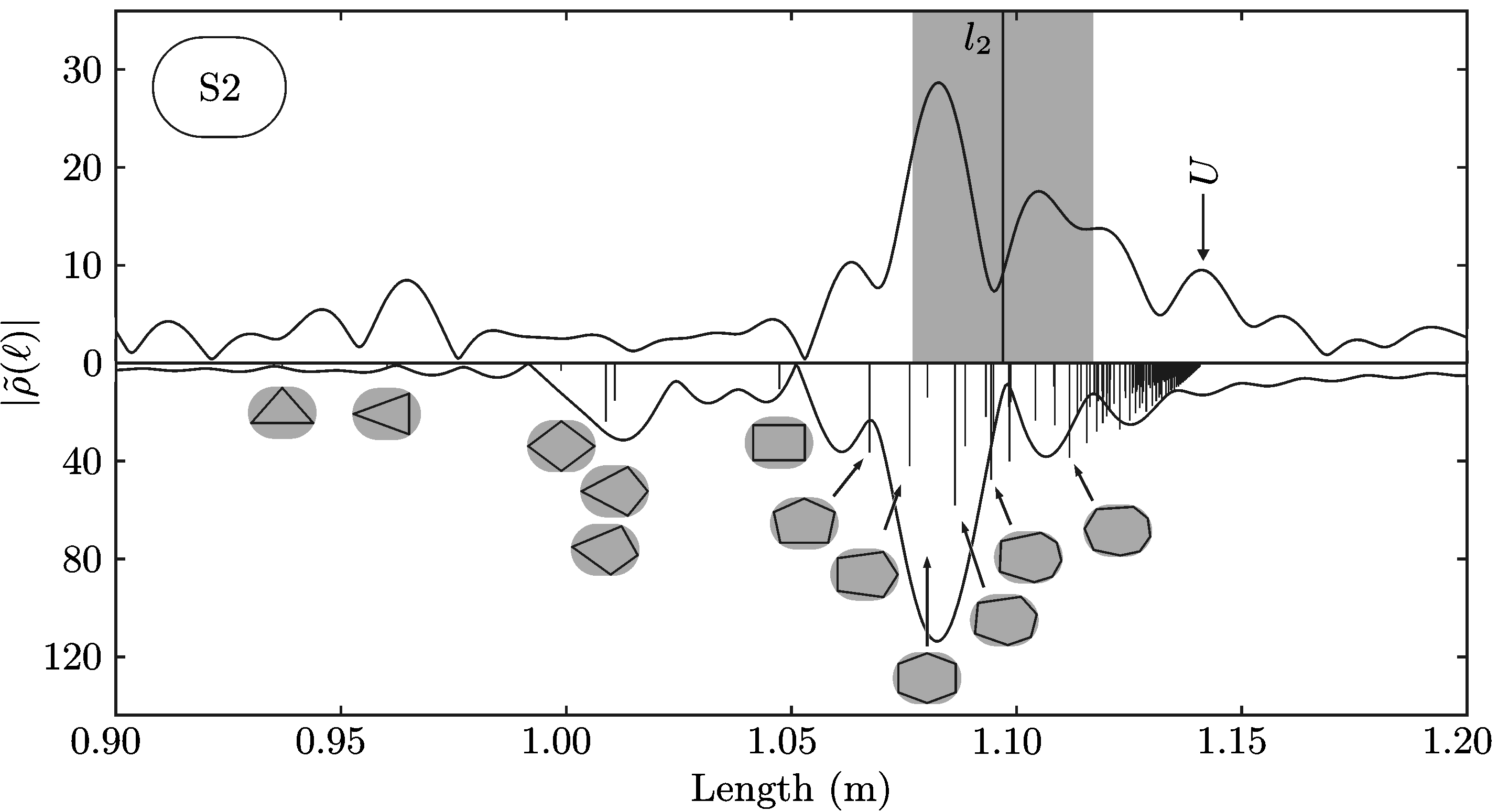}
\end{center}
\caption{\label{fig:lspectS2}Length spectrum for the stadium S2. The top graph shows the experimental length spectrum and the bottom graph the FT of the trace formula. Note the different scales of the top and bottom graphs. The vertical line in the top graph indicates the path length $\lchar_2$, the gray bar its error, and the arrow the circumference $U$ of the stadium. The vertical lines in the bottom graph indicate the lengths and the amplitudes $A_{\ppo, r}$ of the POs. Some POs are shown as insets.}
\end{figure*}

Figures \ref{fig:lspectS1} and \ref{fig:lspectS2} show the length spectra for the stadia S1 and S2, respectively. In each figure the top graph shows the experimental length spectrum. Both are quite similar on a qualitative level and are therefore treated in parallel here. Both length spectra feature two major peaks, which are located at $\ell = 1.262$ m and $1.285$ m in the case of S1 and at $\ell = 1.083$ m and $1.104$ m in the case of S2. Most other peaks are not higher than the oscillating background, which is as large as $|\rhot(\ell)| \approx 5$ in both cases. The bottom graphs of Figs.~\ref{fig:lspectS1} and \ref{fig:lspectS2} show the FT of the corresponding trace formula. The vertical lines in the bottom graphs indicate the lengths $\lpo$ and the modulus of the amplitudes,
\begin{equation} \label{eq:poAmp} A_{\ppo, r} = s_{\ppo} \, |a_{\ppo, r}| \, , \end{equation}
of the POs used in the calculation of the trace formulas. All POs with up to $12$ reflections at the boundary were used. The POs are increasingly dense for lengths close to the circumference $U$, but their amplitudes drop fast for $\lpo \rightarrow U$ because they become more and more unstable. Several families of these whispering gallery type POs with up to $50$ reflections were considered additionally. A full, systematic analysis of these POs and their contributions \cite{Tanner1997}, however, was forgone since the contributions to the trace formula that are relevant in the present work arise from POs with lengths smaller than those of the whispering gallery orbits. The overall shape of the experimental length spectra and the FTs of the trace formula agree well. The vertical lines in the bottom graphs indicate the lengths of the POs. They clearly show that the major peaks in the length spectra in fact result from the interfering contributions of several POs with similar amplitude $A_{\ppo, r}$ each. According to \refeq{eq:singlePOcontr}, the contribution of a single PO has a width $\propto 1 / \Dk$. With $\fmin = 5$ GHz and $\fmax = 20$ GHz here, $1 / \Dk = 3.2$ mm, which is larger than the typical length difference of neighboring POs in the considered length regime. A much larger frequency interval would be needed to resolve the POs; however, we are limited by $f_{2D}$ in the experiment. \newline
Though the shapes of the experimental length spectra and the trace formula predictions agree well for the major peaks, the peak amplitudes of the former are about four times smaller than those of the latter. For the most part, this can be attributed to the large number of missing resonances in the measured frequency spectra. Altogether $137$ resonances in the range of $5$--$20$ GHz were identified for stadium S1, and $138$ resonances in the same frequency range were identified for S2. The total number of resonances in a given frequency interval can be estimated from the Weyl formula \cite{Bogomolny2008},
\begin{equation} \label{eq:Nweyl} \Nweyl(k)= \frac{A n^2}{4 \pi} k^2 + \tilde{r}(n) \frac{U}{4 \pi} k + O(1) \, , \end{equation}
where $\Nweyl(k)$ is the number of resonances up to wave number $k$, $A$ is the area, and $U$ is the circumference of the resonator. The term $\tilde{r}(n)$ accounts for the boundary conditions at the dielectric interface and is given by Eqs.~(28) and (29) in Ref.~\cite{Bogomolny2008}. It can be expressed as \cite{Decanini2003}
\begin{equation} \tilde{r}(n) = \frac{4 n}{\pi} E \left( \frac{n^2 - 1}{n^2} \right) - n  \, , \end{equation}
where $E(x)$ is the complete elliptic integral of the second kind as defined in Ref.~\cite{Abramowitz1972}. The numerical value is $\tilde{r} \approx 1.02$ for the Teflon stadia considered here. According to \refeq{eq:Nweyl}, $3507$ resonances are expected to exist in the regime of $5$--$20$ GHz for stadium S1 and $2643$ for S2, so only $3.9\%$ and $5.2\%$, respectively, of all resonances were identified in the measured spectra. Therefore, investigations of the statistical properties of the measured resonance frequencies and widths are practically impossible. The reason for the large number of unobservable resonances is their short lifetime due to the large radiation losses. A further possible reason for the discrepancy between the experimental and semiclassical peak amplitudes is that corrections to the Fresnel coefficients for reflections at curved interfaces are needed \cite{Hentschel2002}. This is further investigated in \refsec{sec:numLspekt}. \newline
A closer inspection of Figs.~\ref{fig:lspectS1} and \ref{fig:lspectS2} reveals that the POs contributing to the major peaks of the length spectra are all confined by total internal reflection (TIR), i.e., the corresponding angles of incidence at the boundary are all larger than the critical angle for TIR, $\alphacrit = \arcsin{(1 / n)} \approx 45^\circ$. The trace formula also predicts peaks corresponding to POs not confined by TIR, e.g., close to $1.20$ m in \reffig{fig:lspectS1} and close to $1.0$ m in \reffig{fig:lspectS2}, but the experimental length spectra only feature an oscillating background and no real peaks in these length regimes. This is also the case for the BBOs: no significant peaks are seen at $0.6$ m (primitive BBOs, not shown here) or $1.2$ m (first repetition). Summing up, the comparison between the experimental data and the trace formula for the two chaotic stadia yields similar results as in Ref.~\cite{Bittner2010}, where regular dielectric microwave resonators were investigated: there is a good qualitative agreement between the experimental length spectra and the trace formula predictions, the peak amplitudes of the former are smaller than those of the latter due to a large number of missing resonances, and the experimental length spectra show no peaks corresponding to POs not confined by TIR. \newline
The vertical lines and the gray bars in the top graphs of Figs.~\ref{fig:lspectS1} and \ref{fig:lspectS2} indicate the path length corresponding to the subset of equidistant resonances [\refeq{eq:lchar}] and its error, respectively. For both stadia, $\lchar$ is close to the lengths of POs that are confined by TIR, which accounts for the relatively high quality factors of the equidistant resonances. However, $\lchar$ cannot be identified with a single, specific PO, since there are several POs with similar amplitudes $A_{\ppo, r}$ within the error bars of $\lchar$. Therefore, it cannot be clarified whether the set of equidistant resonances can be attributed to scarred states or not. In fact, in Ref.~\cite{Bittner2010} it was found that there is not necessarily a connection between a family of equidistant resonances and a single PO, and in Refs.~\cite{Fang2005a, Fang2007, Fang2007a}, resonant states of dielectric stadia were investigated and scarred states localized on several POs each were found. So the sequences of equidistant resonances in the spectra of S1 and S2 might also be localized on a set of POs, but further investigations are necessary to decide this.

\section{\label{sec:numLspekt}Comparison with numerical calculations}

\begin{figure}[bt]
\begin{center}
\includegraphics[width = 8.4 cm]{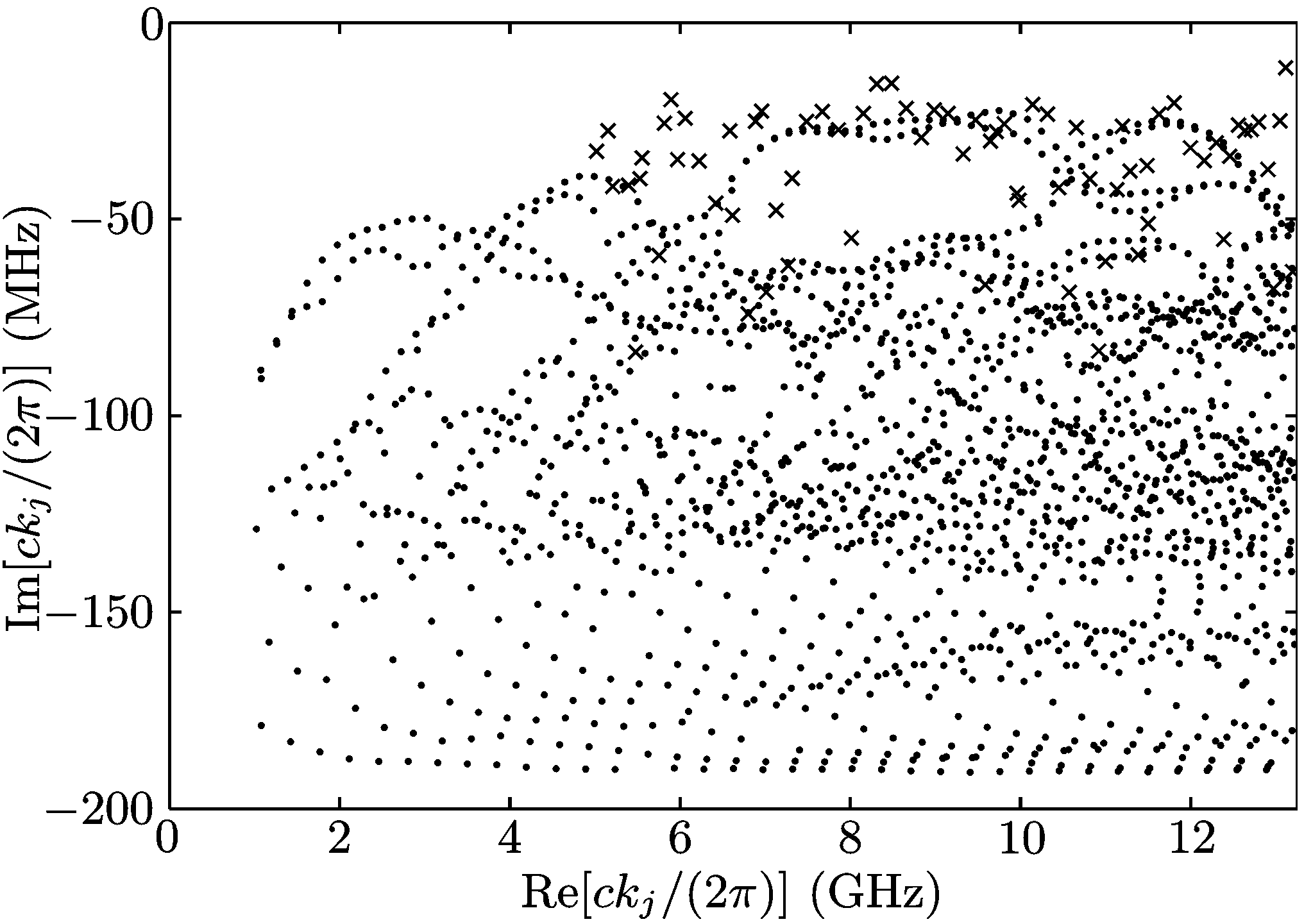}
\end{center}
\caption{\label{fig:reVsIm}Real vs imaginary parts of the eigenvalues for stadium S1 in frequency units. The $\times$ marks are the experimental data, and the dots the numerically calculated data.}
\end{figure}

\begin{figure*}[tb]
\begin{center}
\includegraphics[width = 12 cm]{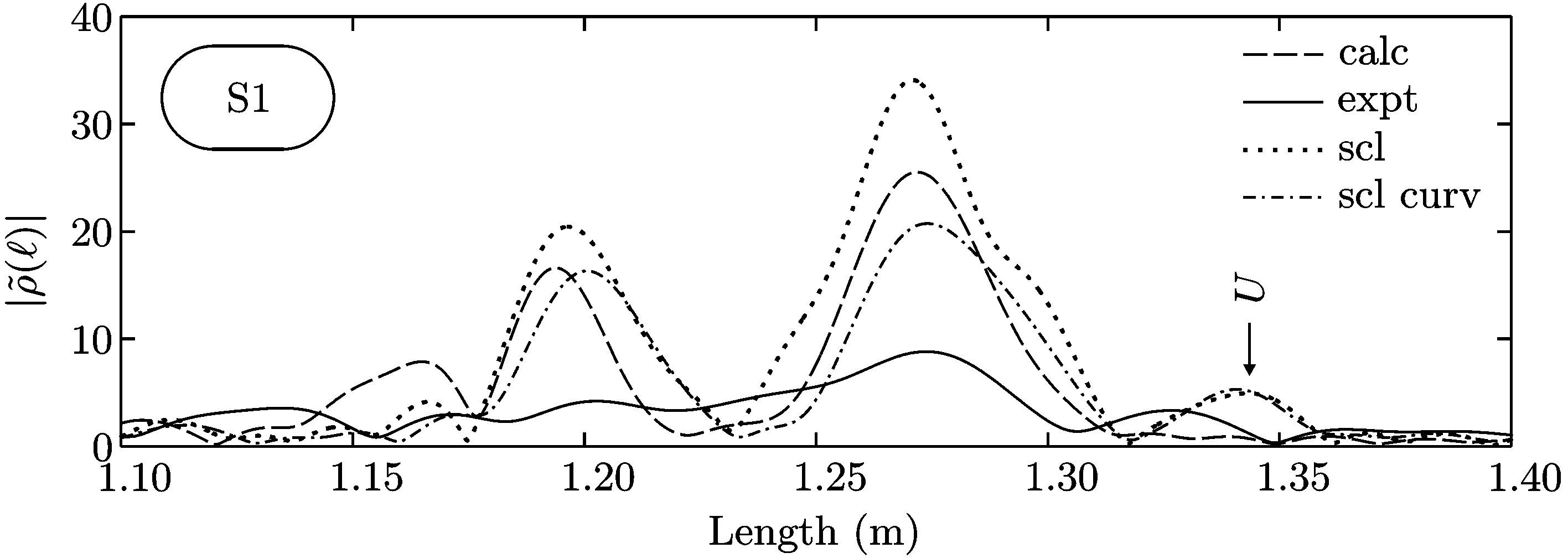}
\end{center}
\caption{\label{fig:lspectS1num}Length spectrum for the stadium S1 evaluated in the frequency range $f = 1.0$--$13.2$ GHz. Shown are the length spectrum of the numerically calculated eigenvalues (dashed line), the experimental one (solid line), the FT of the trace formula with the ordinary Fresnel reflection coefficients [\refeq{eq:fresnelPlan}] inserted (dotted line), and that obtained with the Fresnel reflection coefficients for curved interfaces, \refeq{eq:fresnelCurv} (dash-dotted line). The arrow indicates the circumference $U$.}
\end{figure*}

\begin{figure}[tb]
\begin{center}
\includegraphics[width = 8.4 cm]{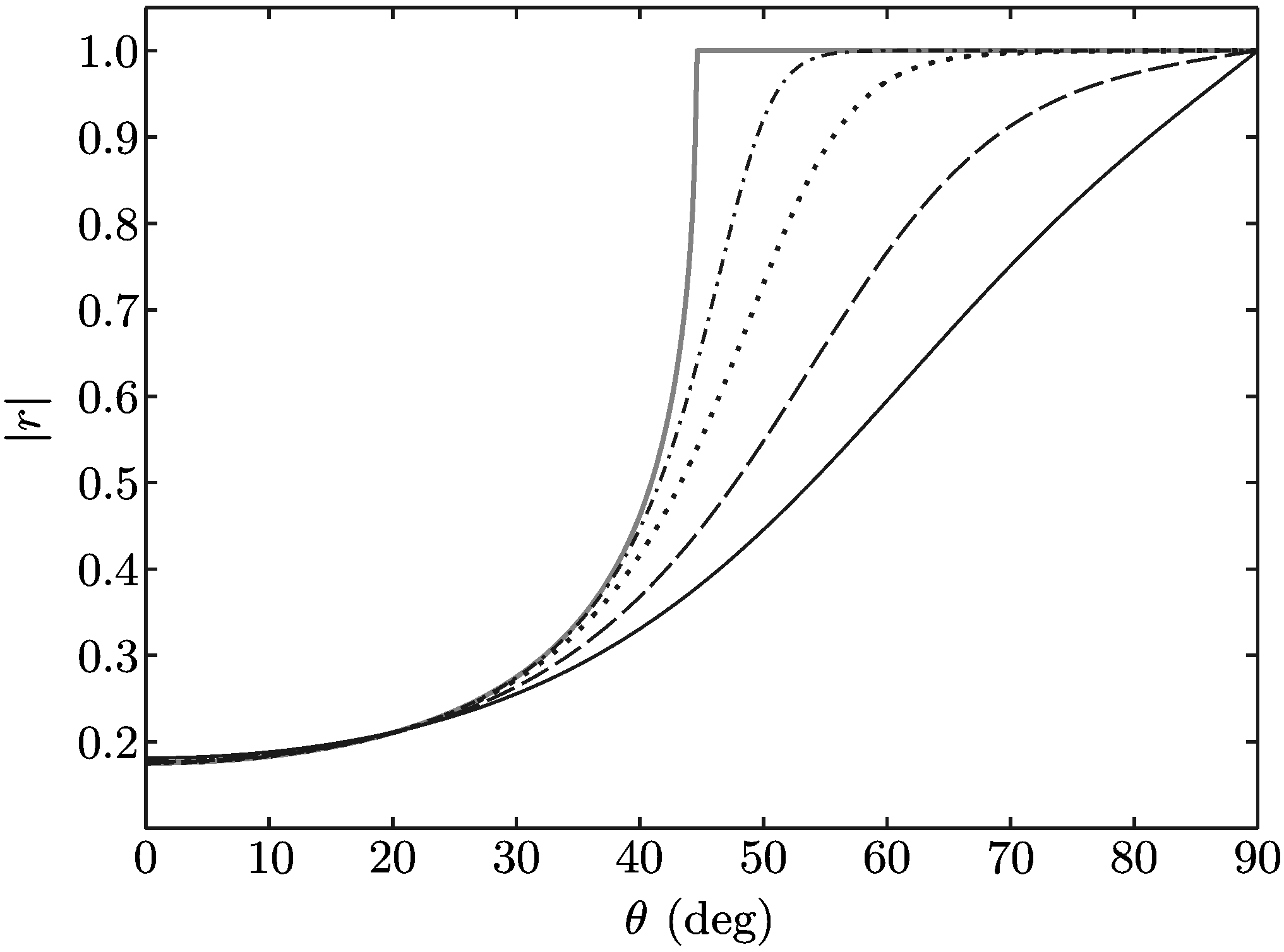}
\end{center}
\caption{\label{fig:fresnel}Modulus of the Fresnel reflection coefficient with respect to the angle of incidence $\theta$ for $n = 1.425$. Shown is the modulus of the reflection coefficient including a curvature correction, $|r_\mathrm{curv}|$, according to \refeq{eq:fresnelCurv} for $kR = 3.1$ corresponding to $f = 1$ GHz with $R = 0.15$ m (solid line), for $kR = 6.3$ ($2$ GHz, dashed line), for $kR = 15.7$ ($5$ GHz, dotted line), and for $kR = 40.9$ ($13$ GHz, dash-dotted line). The gray solid line shows the modulus of the ordinary Fresnel coefficient, $|r|$, defined in \refeq{eq:fresnelPlan}.}
\end{figure}

Since only a small part of all resonances can be identified from measured frequency spectra, we solved the Helmholtz equation [\refeq{eq:helmholtz}] for stadium S1 numerically using a boundary element method \cite{Wiersig2002a} also used, e.g., in Refs.~\cite{Lebental2007, Bogomolny2011}. In the range of $1.0$--$13.2$ GHz a total of $1648$ eigenvalues $k_j$ was found. This is in good agreement with the number of states predicted by Weyl's law, \refeq{eq:Nweyl}. Only $72$ resonances were detected experimentally in the same frequency range, compared to the $137$ resonances found up to $20$ GHz. A survey of the computed eigenvalues is shown in \reffig{fig:reVsIm}, where the imaginary parts of the computed $k_j$ are plotted with respect to the real parts as dots. The resonance frequencies and widths extracted from the measured frequency spectrum are indicated as $\times$ marks. They cover only a narrow strip of the complex plane close to the real axis, i.e., only modes with relatively high quality factors were found experimentally. The calculated and measured eigenvalues agree only roughly, which is attributed to the uncertainties in the determination of the resonance widths and the index of refraction. The length spectrum for the computed eigenvalues is shown in \reffig{fig:lspectS1num} as a dashed line and the experimental length spectrum as a solid line. They look different from those in \reffig{fig:lspectS1} because a smaller frequency/wave number range $\Dk$ is considered here, leading to broader peaks [cf.~\refeq{eq:singlePOcontr}]. The experimental length spectrum shows peaks with amplitudes about three times smaller than that of the numerical one due to the small number of measured resonances. The dotted line is the FT of the trace formula for stadium S1. It shows good qualitative agreement with the numerical length spectrum, even though its peak amplitudes are somewhat larger. This deviation cannot be explained by missing resonances, since according to the Weyl formula only a few resonances, if any, are missing in the set of numerically calculated resonances. A possible explanation is that the trace formula is not very accurate for POs with an incidence angle close to the critical one and that, thus, higher order corrections are needed for such POs \cite{Bogomolny2008}. It is known that the Fresnel reflection coefficients at curved interfaces must be modified depending on the ratio of the radius of curvature $R$ and the wavelength \cite{Hentschel2002}, and it was suggested in Ref.~\cite{Bogomolny2011} to replace the ordinary Fresnel reflection coefficients [\refeq{eq:fresnelPlan}] in the trace formula by ones with a curvature correction, 
\begin{equation} \label{eq:fresnelCurv} r_{j, \mathrm{curv}}(\theta_j, kR) = \frac{n \cos{\theta_j} + i \frac{\hankelp{m}{1}{kR}}{\hankel{m}{1}{kR}}}{n \cos{\theta_j} - i \frac{\hankelp{m}{1}{kR}}{\hankel{m}{1}{kR}}} \, , \end{equation}
where $m = n k R \sin{\theta_j}$, $\hankel{m}{1}{z}$ is a Hankel function of the first kind of order $m$, and $\hankelp{m}{1}{z}$ is its derivative with respect to the argument. The modulus of $r_{j, \mathrm{curv}}$ with respect to the angle of incidence $\theta$ for different values of $kR$ is compared to that of the ordinary Fresnel coefficient in \reffig{fig:fresnel}. The deviations between the corrected and the ordinary Fresnel coefficients is considerable, especially close to the critical angle and for low frequencies, that is, small $kR$. The ordinary Fresnel coefficients are recovered in the limit $kR \rightarrow \infty$. The curvature-corrected Fresnel coefficients were inserted into the trace formula, \refeq{eq:trFormChaot}, and its FT was calculated numerically to obtain the dash-dotted curve in \reffig{fig:lspectS1num}. As expected, its peak amplitudes are smaller than those obtained for the trace formula without including curvature corrections. In the case of the peak near $1.20$ m, the peak amplitude matches that of the numerical length spectrum, and in the case of the peak near $1.27$ m, it is even a bit smaller. Furthermore, near $1.20$ m the peak of the trace formula with curvature correction is shifted to the right with respect to that of the ordinary trace formula. This can be explained by the frequency dependence of the phase of the curvature-corrected Fresnel coefficients \cite{Bittner2012a}. In summary, the introduction of curvature corrections leads to a better agreement between the calculated length spectrum and the trace formula predictions; however, deviations remain. Since the major peaks in the length spectrum result from the contributions of several POs each (cf.~\reffig{fig:lspectS1}), a detailed analysis of the influence of the curvature correction on the contribution of a single PO is not possible here. This is necessary also because corrections to the trace formula apply for POs that are not well isolated \cite{Takami1995, Schomerus1997}. Therefore, further investigations of higher-order corrections to the semiclassical trace formula are needed.

\section{\label{sec:concl}Conclusions}

The frequency spectra of two passive 2D dielectric microwave resonators with chaotic classical dynamics were measured and analyzed. The corresponding length spectra were compared to the predictions of a semiclassical trace formula proposed in Ref.~\cite{Bogomolny2008}. Good qualitative agreement between the experimental length spectra and the predictions of the trace formula was found. However, the peak amplitudes of the experimental length spectra were systematically smaller than those predicted by the trace formula, which is mostly attributed to the large number of missing resonances in the measured spectra. Furthermore, no contributions from POs that are not confined by TIR were found in the experimental length spectra, even if predicted by the trace formula. Apparently, the experimentally observable, long-lived resonances correspond to the best-confined POs of the corresponding billiard, even though individual resonances cannot be associated with specific trajectories like for systems with regular classical dynamics, e.g., the circle or square \cite{Bittner2010}. The same observations concerning the amplitudes of and the POs contributing to the experimental length spectra were made in experiments with regular dielectric resonators in Ref.~\cite{Bittner2010}, so they seem to be valid for all types of dielectric resonators. Furthermore, the measured frequency spectra contained subsets of equidistant resonances, and we investigated whether these resonant modes are localized on a particular PO. There are indications that they are instead localized on several POs each, but no clear conclusion could be drawn from the data. Spectra containing one or more sequences of equidistant resonances were also observed for other dielectric resonators with low index of refraction ($n \approx 1.3$--$1.5$) and various shapes \cite{Lebental2006, Djellali2009, Bogomolny2011, Bittner2010, Lee2009, Chern2004}. Thus, this seems to be a generic phenomenon for dielectric resonators with low $n$; however, the origin and whether it is a common feature is not known. Finally, we computed the resonances of one of the investigated resonators numerically to obtain a complete spectrum of eigenvalues. The corresponding length spectrum has been compared to the corresponding experimental one and to the trace formula prediction. It was concluded that the trace formula overestimates the peak amplitudes of the length spectrum. We applied a curvature correction to the Fresnel reflection coefficients entering the trace formula and found better agreement between the numerical length spectrum and the trace formula prediction. However, some deviations remained, and it would be worthwhile to systematically investigate further corrections to the trace formula of Ref.~\cite{Bogomolny2008} for a more complete understanding of the length spectra of dielectric resonators. Corrections to the Fresnel reflection coefficients for finite wavelengths are of general interest for the understanding of the ray-wave correspondence in dielectric resonators (see, for example Refs.~\cite{Hentschel2002, Altmann2008c, Unterhinninghofen2010a}).

\begin{acknowledgments}
We thank E. Bogomolny for intense discussions and hospitality at the LPTMS in Orsay. The numerical calculations of resonance frequencies are based on a code developed by C. Schmit. This work was supported by the DFG within the Sonderforschungsbereich 634.
\end{acknowledgments}


\begin{thebibliography}{71}%
\makeatletter
\providecommand \@ifxundefined [1]{%
 \@ifx{#1\undefined}
}%
\providecommand \@ifnum [1]{%
 \ifnum #1\expandafter \@firstoftwo
 \else \expandafter \@secondoftwo
 \fi
}%
\providecommand \@ifx [1]{%
 \ifx #1\expandafter \@firstoftwo
 \else \expandafter \@secondoftwo
 \fi
}%
\providecommand \natexlab [1]{#1}%
\providecommand \enquote  [1]{``#1''}%
\providecommand \bibnamefont  [1]{#1}%
\providecommand \bibfnamefont [1]{#1}%
\providecommand \citenamefont [1]{#1}%
\providecommand \href@noop [0]{\@secondoftwo}%
\providecommand \href [0]{\begingroup \@sanitize@url \@href}%
\providecommand \@href[1]{\@@startlink{#1}\@@href}%
\providecommand \@@href[1]{\endgroup#1\@@endlink}%
\providecommand \@sanitize@url [0]{\catcode `\\12\catcode `\$12\catcode
  `\&12\catcode `\#12\catcode `\^12\catcode `\_12\catcode `\%12\relax}%
\providecommand \@@startlink[1]{}%
\providecommand \@@endlink[0]{}%
\providecommand \url  [0]{\begingroup\@sanitize@url \@url }%
\providecommand \@url [1]{\endgroup\@href {#1}{\urlprefix }}%
\providecommand \urlprefix  [0]{URL }%
\providecommand \Eprint [0]{\href }%
\providecommand \doibase [0]{http://dx.doi.org/}%
\providecommand \selectlanguage [0]{\@gobble}%
\providecommand \bibinfo  [0]{\@secondoftwo}%
\providecommand \bibfield  [0]{\@secondoftwo}%
\providecommand \translation [1]{[#1]}%
\providecommand \BibitemOpen [0]{}%
\providecommand \bibitemStop [0]{}%
\providecommand \bibitemNoStop [0]{.\EOS\space}%
\providecommand \EOS [0]{\spacefactor3000\relax}%
\providecommand \BibitemShut  [1]{\csname bibitem#1\endcsname}%
\let\auto@bib@innerbib\@empty
\bibitem [{\citenamefont {N\"ockel}\ and\ \citenamefont
  {Chang}(2003)}]{Nockel2003}%
  \BibitemOpen
  \bibfield  {author} {\bibinfo {author} {\bibfnamefont {J.~U.}\ \bibnamefont
  {N\"ockel}}\ and\ \bibinfo {author} {\bibfnamefont {R.~K.}\ \bibnamefont
  {Chang}},\ }\enquote {\bibinfo {title} {Cavity-enhanced spectroscopies},}\ \
  (\bibinfo  {publisher} {Academic Press},\ \bibinfo {address} {San Diego},\
  \bibinfo {year} {2003})\ Chap.\ \bibinfo {chapter} {6 "2D microcavities:
  Theory and experiments"}, pp.\ \bibinfo {pages} {185–--226}\BibitemShut
  {NoStop}%
\bibitem [{\citenamefont {Vahala}(2004)}]{Vahala2004}%
  \BibitemOpen
  \bibinfo {editor} {\bibfnamefont {K.}~\bibnamefont {Vahala}},\ ed.,\
  \href@noop {} {\emph {\bibinfo {title} {Optical Microcavities}}}\ (\bibinfo
  {publisher} {World Scientific},\ \bibinfo {address} {Singapore},\ \bibinfo
  {year} {2004})\BibitemShut {NoStop}%
\bibitem [{\citenamefont {Matsko}\ \emph {et~al.}(2005)\citenamefont {Matsko},
  \citenamefont {Savchenkov}, \citenamefont {Strekalov}, \citenamefont
  {Ilchenko},\ and\ \citenamefont {Maleki}}]{Matsko2005}%
  \BibitemOpen
  \bibfield  {author} {\bibinfo {author} {\bibfnamefont {A.~B.}\ \bibnamefont
  {Matsko}}, \bibinfo {author} {\bibfnamefont {A.~A.}\ \bibnamefont
  {Savchenkov}}, \bibinfo {author} {\bibfnamefont {D.}~\bibnamefont
  {Strekalov}}, \bibinfo {author} {\bibfnamefont {V.~S.}\ \bibnamefont
  {Ilchenko}}, \ and\ \bibinfo {author} {\bibfnamefont {L.}~\bibnamefont
  {Maleki}},\ }\href@noop {} {}\bibinfo {howpublished} {IPN Progress Report
  42-162} (\bibinfo {year} {2005})\BibitemShut {NoStop}%
\bibitem [{\citenamefont {McCall}\ \emph {et~al.}(1992)\citenamefont {McCall},
  \citenamefont {Levi}, \citenamefont {Slusher}, \citenamefont {Pearton},\ and\
  \citenamefont {Logan}}]{McCall1992}%
  \BibitemOpen
  \bibfield  {author} {\bibinfo {author} {\bibfnamefont {S.~L.}\ \bibnamefont
  {McCall}}, \bibinfo {author} {\bibfnamefont {A.~F.~J.}\ \bibnamefont {Levi}},
  \bibinfo {author} {\bibfnamefont {R.~E.}\ \bibnamefont {Slusher}}, \bibinfo
  {author} {\bibfnamefont {S.~J.}\ \bibnamefont {Pearton}}, \ and\ \bibinfo
  {author} {\bibfnamefont {R.~A.}\ \bibnamefont {Logan}},\ }\href {\doibase
  DOI:10.1063/1.106688} {\bibfield  {journal} {\bibinfo  {journal} {Appl. Phys.
  Lett.}\ }\textbf {\bibinfo {volume} {60}},\ \bibinfo {pages} {289} (\bibinfo
  {year} {1992})}\BibitemShut {NoStop}%
\bibitem [{\citenamefont {N{\"o}ckel}\ \emph {et~al.}(1994)\citenamefont
  {N{\"o}ckel}, \citenamefont {Stone},\ and\ \citenamefont
  {Chang}}]{Noeckel1994}%
  \BibitemOpen
  \bibfield  {author} {\bibinfo {author} {\bibfnamefont {J.~U.}\ \bibnamefont
  {N{\"o}ckel}}, \bibinfo {author} {\bibfnamefont {A.~D.}\ \bibnamefont
  {Stone}}, \ and\ \bibinfo {author} {\bibfnamefont {R.~K.}\ \bibnamefont
  {Chang}},\ }\href {\doibase 10.1364/OL.19.001693} {\bibfield  {journal}
  {\bibinfo  {journal} {Opt. Lett.}\ }\textbf {\bibinfo {volume} {19}},\
  \bibinfo {pages} {1693} (\bibinfo {year} {1994})}\BibitemShut {NoStop}%
\bibitem [{\citenamefont {N{\"o}ckel}\ and\ \citenamefont
  {Stone}(1997)}]{Noeckel1997}%
  \BibitemOpen
  \bibfield  {author} {\bibinfo {author} {\bibfnamefont {J.~U.}\ \bibnamefont
  {N{\"o}ckel}}\ and\ \bibinfo {author} {\bibfnamefont {A.~D.}\ \bibnamefont
  {Stone}},\ }\href {\doibase 10.1038/385045a0} {\bibfield  {journal} {\bibinfo
   {journal} {Nature (London)}\ }\textbf {\bibinfo {volume} {385}},\ \bibinfo
  {pages} {45} (\bibinfo {year} {1997})}\BibitemShut {NoStop}%
\bibitem [{\citenamefont {Altmann}(2009)}]{Altmann2009}%
  \BibitemOpen
  \bibfield  {author} {\bibinfo {author} {\bibfnamefont {E.~G.}\ \bibnamefont
  {Altmann}},\ }\href {\doibase 10.1103/PhysRevA.79.013830} {\bibfield
  {journal} {\bibinfo  {journal} {Phys. Rev. A}\ }\textbf {\bibinfo {volume}
  {79}},\ \bibinfo {pages} {013830} (\bibinfo {year} {2009})}\BibitemShut
  {NoStop}%
\bibitem [{\citenamefont {Rex}\ \emph {et~al.}(2002)\citenamefont {Rex},
  \citenamefont {Tureci}, \citenamefont {Schwefel}, \citenamefont {Chang},\
  and\ \citenamefont {Stone}}]{Rex2002}%
  \BibitemOpen
  \bibfield  {author} {\bibinfo {author} {\bibfnamefont {N.~B.}\ \bibnamefont
  {Rex}}, \bibinfo {author} {\bibfnamefont {H.~E.}\ \bibnamefont {Tureci}},
  \bibinfo {author} {\bibfnamefont {H.~G.~L.}\ \bibnamefont {Schwefel}},
  \bibinfo {author} {\bibfnamefont {R.~K.}\ \bibnamefont {Chang}}, \ and\
  \bibinfo {author} {\bibfnamefont {A.~D.}\ \bibnamefont {Stone}},\ }\href
  {\doibase 10.1103/PhysRevLett.88.094102} {\bibfield  {journal} {\bibinfo
  {journal} {Phys. Rev. Lett.}\ }\textbf {\bibinfo {volume} {88}},\ \bibinfo
  {pages} {094102} (\bibinfo {year} {2002})}\BibitemShut {NoStop}%
\bibitem [{\citenamefont {Gmachl}\ \emph {et~al.}(2002)\citenamefont {Gmachl},
  \citenamefont {Narimanov}, \citenamefont {Capasso}, \citenamefont
  {Baillargeon},\ and\ \citenamefont {Cho}}]{Gmachl2002}%
  \BibitemOpen
  \bibfield  {author} {\bibinfo {author} {\bibfnamefont {C.}~\bibnamefont
  {Gmachl}}, \bibinfo {author} {\bibfnamefont {E.~E.}\ \bibnamefont
  {Narimanov}}, \bibinfo {author} {\bibfnamefont {F.}~\bibnamefont {Capasso}},
  \bibinfo {author} {\bibfnamefont {J.~N.}\ \bibnamefont {Baillargeon}}, \ and\
  \bibinfo {author} {\bibfnamefont {A.~Y.}\ \bibnamefont {Cho}},\ }\href
  {\doibase 10.1364/OL.27.000824} {\bibfield  {journal} {\bibinfo  {journal}
  {Opt. Lett.}\ }\textbf {\bibinfo {volume} {27}},\ \bibinfo {pages} {824}
  (\bibinfo {year} {2002})}\BibitemShut {NoStop}%
\bibitem [{\citenamefont {Harayama}\ \emph {et~al.}(2003)\citenamefont
  {Harayama}, \citenamefont {Fukushima}, \citenamefont {Davis}, \citenamefont
  {Vaccaro}, \citenamefont {Miyasaka}, \citenamefont {Nishimura},\ and\
  \citenamefont {Aida}}]{Harayama2003}%
  \BibitemOpen
  \bibfield  {author} {\bibinfo {author} {\bibfnamefont {T.}~\bibnamefont
  {Harayama}}, \bibinfo {author} {\bibfnamefont {T.}~\bibnamefont {Fukushima}},
  \bibinfo {author} {\bibfnamefont {P.}~\bibnamefont {Davis}}, \bibinfo
  {author} {\bibfnamefont {P.~O.}\ \bibnamefont {Vaccaro}}, \bibinfo {author}
  {\bibfnamefont {T.}~\bibnamefont {Miyasaka}}, \bibinfo {author}
  {\bibfnamefont {T.}~\bibnamefont {Nishimura}}, \ and\ \bibinfo {author}
  {\bibfnamefont {T.}~\bibnamefont {Aida}},\ }\href {\doibase
  10.1103/PhysRevE.67.015207} {\bibfield  {journal} {\bibinfo  {journal} {Phys.
  Rev. E}\ }\textbf {\bibinfo {volume} {67}},\ \bibinfo {pages} {015207}
  (\bibinfo {year} {2003})}\BibitemShut {NoStop}%
\bibitem [{\citenamefont {Fang}\ \emph {et~al.}(2005)\citenamefont {Fang},
  \citenamefont {Yamilov},\ and\ \citenamefont {Cao}}]{Fang2005a}%
  \BibitemOpen
  \bibfield  {author} {\bibinfo {author} {\bibfnamefont {W.}~\bibnamefont
  {Fang}}, \bibinfo {author} {\bibfnamefont {A.}~\bibnamefont {Yamilov}}, \
  and\ \bibinfo {author} {\bibfnamefont {H.}~\bibnamefont {Cao}},\ }\href
  {\doibase 10.1103/PhysRevA.72.023815} {\bibfield  {journal} {\bibinfo
  {journal} {Phys. Rev. A}\ }\textbf {\bibinfo {volume} {72}},\ \bibinfo
  {pages} {023815} (\bibinfo {year} {2005})}\BibitemShut {NoStop}%
\bibitem [{\citenamefont {Fang}\ \emph {et~al.}(2007)\citenamefont {Fang},
  \citenamefont {Cao},\ and\ \citenamefont {Solomon}}]{Fang2007}%
  \BibitemOpen
  \bibfield  {author} {\bibinfo {author} {\bibfnamefont {W.}~\bibnamefont
  {Fang}}, \bibinfo {author} {\bibfnamefont {H.}~\bibnamefont {Cao}}, \ and\
  \bibinfo {author} {\bibfnamefont {G.~S.}\ \bibnamefont {Solomon}},\ }\href
  {\doibase 10.1063/1.2535692} {\bibfield  {journal} {\bibinfo  {journal}
  {Appl. Phys. Lett.}\ }\textbf {\bibinfo {volume} {90}},\ \bibinfo {pages}
  {081108} (\bibinfo {year} {2007})}\BibitemShut {NoStop}%
\bibitem [{\citenamefont {Fang}\ and\ \citenamefont {Cao}(2007)}]{Fang2007a}%
  \BibitemOpen
  \bibfield  {author} {\bibinfo {author} {\bibfnamefont {W.}~\bibnamefont
  {Fang}}\ and\ \bibinfo {author} {\bibfnamefont {H.}~\bibnamefont {Cao}},\
  }\href {\doibase 10.1063/1.2762285} {\bibfield  {journal} {\bibinfo
  {journal} {Appl. Phys. Lett.}\ }\textbf {\bibinfo {volume} {91}},\ \bibinfo
  {pages} {041108} (\bibinfo {year} {2007})}\BibitemShut {NoStop}%
\bibitem [{\citenamefont {Gutzwiller}(1970)}]{Gutzwiller1970}%
  \BibitemOpen
  \bibfield  {author} {\bibinfo {author} {\bibfnamefont {M.~C.}\ \bibnamefont
  {Gutzwiller}},\ }\href {\doibase 10.1063/1.1665328} {\bibfield  {journal}
  {\bibinfo  {journal} {J. Math. Phys.}\ }\textbf {\bibinfo {volume} {11}},\
  \bibinfo {pages} {1791} (\bibinfo {year} {1970})}\BibitemShut {NoStop}%
\bibitem [{\citenamefont {Gutzwiller}(1971)}]{Gutzwiller1971}%
  \BibitemOpen
  \bibfield  {author} {\bibinfo {author} {\bibfnamefont {M.~C.}\ \bibnamefont
  {Gutzwiller}},\ }\href {\doibase 10.1063/1.1665596} {\bibfield  {journal}
  {\bibinfo  {journal} {J. Math. Phys.}\ }\textbf {\bibinfo {volume} {12}},\
  \bibinfo {pages} {343} (\bibinfo {year} {1971})}\BibitemShut {NoStop}%
\bibitem [{\citenamefont {Brack}\ and\ \citenamefont
  {Bhaduri}(2003)}]{Brack2003}%
  \BibitemOpen
  \bibfield  {author} {\bibinfo {author} {\bibfnamefont {M.}~\bibnamefont
  {Brack}}\ and\ \bibinfo {author} {\bibfnamefont {R.~K.}\ \bibnamefont
  {Bhaduri}},\ }\href@noop {} {\emph {\bibinfo {title} {Semiclassical
  Physics}}}\ (\bibinfo  {publisher} {Westview Press},\ \bibinfo {address}
  {Oxford},\ \bibinfo {year} {2003})\BibitemShut {NoStop}%
\bibitem [{\citenamefont {Bogomolny}\ \emph {et~al.}(2008)\citenamefont
  {Bogomolny}, \citenamefont {Dubertrand},\ and\ \citenamefont
  {Schmit}}]{Bogomolny2008}%
  \BibitemOpen
  \bibfield  {author} {\bibinfo {author} {\bibfnamefont {E.}~\bibnamefont
  {Bogomolny}}, \bibinfo {author} {\bibfnamefont {R.}~\bibnamefont
  {Dubertrand}}, \ and\ \bibinfo {author} {\bibfnamefont {C.}~\bibnamefont
  {Schmit}},\ }\href {\doibase 10.1103/PhysRevE.78.056202} {\bibfield
  {journal} {\bibinfo  {journal} {Phys. Rev. E}\ }\textbf {\bibinfo {volume}
  {78}},\ \bibinfo {pages} {056202} (\bibinfo {year} {2008})}\BibitemShut
  {NoStop}%
\bibitem [{\citenamefont {Hales}\ \emph {et~al.}(2011)\citenamefont {Hales},
  \citenamefont {Sieber},\ and\ \citenamefont {Waalkens}}]{Hales2011}%
  \BibitemOpen
  \bibfield  {author} {\bibinfo {author} {\bibfnamefont {R.~F.~M.}\
  \bibnamefont {Hales}}, \bibinfo {author} {\bibfnamefont {M.}~\bibnamefont
  {Sieber}}, \ and\ \bibinfo {author} {\bibfnamefont {H.}~\bibnamefont
  {Waalkens}},\ }\href {\doibase 10.1088/1751-8113/44/15/155305} {\bibfield
  {journal} {\bibinfo  {journal} {J. Phys. A}\ }\textbf {\bibinfo {volume}
  {44}},\ \bibinfo {pages} {155305} (\bibinfo {year} {2011})}\BibitemShut
  {NoStop}%
\bibitem [{\citenamefont {Lebental}\ \emph {et~al.}(2007)\citenamefont
  {Lebental}, \citenamefont {Djellali}, \citenamefont {Arnaud}, \citenamefont
  {Lauret}, \citenamefont {Zyss}, \citenamefont {Dubertrand}, \citenamefont
  {Schmit},\ and\ \citenamefont {Bogomolny}}]{Lebental2007}%
  \BibitemOpen
  \bibfield  {author} {\bibinfo {author} {\bibfnamefont {M.}~\bibnamefont
  {Lebental}}, \bibinfo {author} {\bibfnamefont {N.}~\bibnamefont {Djellali}},
  \bibinfo {author} {\bibfnamefont {C.}~\bibnamefont {Arnaud}}, \bibinfo
  {author} {\bibfnamefont {J.-S.}\ \bibnamefont {Lauret}}, \bibinfo {author}
  {\bibfnamefont {J.}~\bibnamefont {Zyss}}, \bibinfo {author} {\bibfnamefont
  {R.}~\bibnamefont {Dubertrand}}, \bibinfo {author} {\bibfnamefont
  {C.}~\bibnamefont {Schmit}}, \ and\ \bibinfo {author} {\bibfnamefont
  {E.}~\bibnamefont {Bogomolny}},\ }\href {\doibase 10.1103/PhysRevA.76.023830}
  {\bibfield  {journal} {\bibinfo  {journal} {Phys. Rev. A}\ }\textbf {\bibinfo
  {volume} {76}},\ \bibinfo {pages} {023830} (\bibinfo {year}
  {2007})}\BibitemShut {NoStop}%
\bibitem [{\citenamefont {Bogomolny}\ \emph {et~al.}(2011)\citenamefont
  {Bogomolny}, \citenamefont {Djellali}, \citenamefont {Dubertrand},
  \citenamefont {Gozhyk}, \citenamefont {Lebental}, \citenamefont {Schmit},
  \citenamefont {Ulysse},\ and\ \citenamefont {Zyss}}]{Bogomolny2011}%
  \BibitemOpen
  \bibfield  {author} {\bibinfo {author} {\bibfnamefont {E.}~\bibnamefont
  {Bogomolny}}, \bibinfo {author} {\bibfnamefont {N.}~\bibnamefont {Djellali}},
  \bibinfo {author} {\bibfnamefont {R.}~\bibnamefont {Dubertrand}}, \bibinfo
  {author} {\bibfnamefont {I.}~\bibnamefont {Gozhyk}}, \bibinfo {author}
  {\bibfnamefont {M.}~\bibnamefont {Lebental}}, \bibinfo {author}
  {\bibfnamefont {C.}~\bibnamefont {Schmit}}, \bibinfo {author} {\bibfnamefont
  {C.}~\bibnamefont {Ulysse}}, \ and\ \bibinfo {author} {\bibfnamefont
  {J.}~\bibnamefont {Zyss}},\ }\href {\doibase 10.1103/PhysRevE.83.036208}
  {\bibfield  {journal} {\bibinfo  {journal} {Phys. Rev. E}\ }\textbf {\bibinfo
  {volume} {83}},\ \bibinfo {pages} {036208} (\bibinfo {year}
  {2011})}\BibitemShut {NoStop}%
\bibitem [{\citenamefont {Bittner}\ \emph {et~al.}(2010)\citenamefont
  {Bittner}, \citenamefont {Bogomolny}, \citenamefont {Dietz}, \citenamefont
  {Miski-Oglu}, \citenamefont {Oria~Iriarte}, \citenamefont {Richter},\ and\
  \citenamefont {Sch{\"a}fer}}]{Bittner2010}%
  \BibitemOpen
  \bibfield  {author} {\bibinfo {author} {\bibfnamefont {S.}~\bibnamefont
  {Bittner}}, \bibinfo {author} {\bibfnamefont {E.}~\bibnamefont {Bogomolny}},
  \bibinfo {author} {\bibfnamefont {B.}~\bibnamefont {Dietz}}, \bibinfo
  {author} {\bibfnamefont {M.}~\bibnamefont {Miski-Oglu}}, \bibinfo {author}
  {\bibfnamefont {P.}~\bibnamefont {Oria~Iriarte}}, \bibinfo {author}
  {\bibfnamefont {A.}~\bibnamefont {Richter}}, \ and\ \bibinfo {author}
  {\bibfnamefont {F.}~\bibnamefont {Sch{\"a}fer}},\ }\href {\doibase
  10.1103/PhysRevE.81.066215} {\bibfield  {journal} {\bibinfo  {journal} {Phys.
  Rev. E}\ }\textbf {\bibinfo {volume} {81}},\ \bibinfo {pages} {066215}
  (\bibinfo {year} {2010})}\BibitemShut {NoStop}%
\bibitem [{\citenamefont {Bittner}\ \emph {et~al.}(2011)\citenamefont
  {Bittner}, \citenamefont {Dietz},\ and\ \citenamefont
  {Richter}}]{Bittner2011a}%
  \BibitemOpen
  \bibfield  {author} {\bibinfo {author} {\bibfnamefont {S.}~\bibnamefont
  {Bittner}}, \bibinfo {author} {\bibfnamefont {B.}~\bibnamefont {Dietz}}, \
  and\ \bibinfo {author} {\bibfnamefont {A.}~\bibnamefont {Richter}},\
  }\enquote {\bibinfo {title} {Trends in nano- and micro-cavities},}\ \
  (\bibinfo  {publisher} {Bentham Science Publishers},\ \bibinfo {year}
  {2011})\ Chap.\ \bibinfo {chapter} {1 "Semiclassical approaches for
  dielectric resonators"}, pp.\ \bibinfo {pages} {1--39}\BibitemShut {NoStop}%
\bibitem [{\citenamefont {Bittner}\ \emph {et~al.}(2012)\citenamefont
  {Bittner}, \citenamefont {Bogomolny}, \citenamefont {Dietz}, \citenamefont
  {Miski-Oglu},\ and\ \citenamefont {Richter}}]{Bittner2012a}%
  \BibitemOpen
  \bibfield  {author} {\bibinfo {author} {\bibfnamefont {S.}~\bibnamefont
  {Bittner}}, \bibinfo {author} {\bibfnamefont {E.}~\bibnamefont {Bogomolny}},
  \bibinfo {author} {\bibfnamefont {B.}~\bibnamefont {Dietz}}, \bibinfo
  {author} {\bibfnamefont {M.}~\bibnamefont {Miski-Oglu}}, \ and\ \bibinfo
  {author} {\bibfnamefont {A.}~\bibnamefont {Richter}},\ }\href {\doibase
  10.1103/PhysRevE.85.026203} {\bibfield  {journal} {\bibinfo  {journal} {Phys.
  Rev. E}\ }\textbf {\bibinfo {volume} {85}},\ \bibinfo {pages} {026203}
  (\bibinfo {year} {2012})}\BibitemShut {NoStop}%
\bibitem [{\citenamefont {Bunimovich}(1979)}]{Bunimovich1979}%
  \BibitemOpen
  \bibfield  {author} {\bibinfo {author} {\bibfnamefont {L.~A.}\ \bibnamefont
  {Bunimovich}},\ }\href@noop {} {\bibfield  {journal} {\bibinfo  {journal}
  {Commun. Math. Phys.}\ }\textbf {\bibinfo {volume} {65}},\ \bibinfo {pages}
  {295} (\bibinfo {year} {1979})}\BibitemShut {NoStop}%
\bibitem [{\citenamefont {McDonald}\ and\ \citenamefont
  {Kaufman}(1979)}]{McDonald1979}%
  \BibitemOpen
  \bibfield  {author} {\bibinfo {author} {\bibfnamefont {S.~W.}\ \bibnamefont
  {McDonald}}\ and\ \bibinfo {author} {\bibfnamefont {A.~N.}\ \bibnamefont
  {Kaufman}},\ }\href {\doibase 10.1103/PhysRevLett.42.1189} {\bibfield
  {journal} {\bibinfo  {journal} {Phys. Rev. Lett.}\ }\textbf {\bibinfo
  {volume} {42}},\ \bibinfo {pages} {1189} (\bibinfo {year}
  {1979})}\BibitemShut {NoStop}%
\bibitem [{\citenamefont {Christoffel}\ and\ \citenamefont
  {Brumer}(1986)}]{Christoffel1986}%
  \BibitemOpen
  \bibfield  {author} {\bibinfo {author} {\bibfnamefont {K.~M.}\ \bibnamefont
  {Christoffel}}\ and\ \bibinfo {author} {\bibfnamefont {P.}~\bibnamefont
  {Brumer}},\ }\href {\doibase 10.1103/PhysRevA.33.1309} {\bibfield  {journal}
  {\bibinfo  {journal} {Phys. Rev. A}\ }\textbf {\bibinfo {volume} {33}},\
  \bibinfo {pages} {1309} (\bibinfo {year} {1986})}\BibitemShut {NoStop}%
\bibitem [{\citenamefont {Bogomolny}(1988)}]{Bogomolny1988}%
  \BibitemOpen
  \bibfield  {author} {\bibinfo {author} {\bibfnamefont {E.}~\bibnamefont
  {Bogomolny}},\ }\href {\doibase 10.1016/0167-2789(88)90075-9} {\bibfield
  {journal} {\bibinfo  {journal} {Physica D}\ }\textbf {\bibinfo {volume}
  {31}},\ \bibinfo {pages} {169} (\bibinfo {year} {1988})}\BibitemShut
  {NoStop}%
\bibitem [{\citenamefont {Shudo}\ and\ \citenamefont
  {Shimizu}(1990)}]{Shudo1990}%
  \BibitemOpen
  \bibfield  {author} {\bibinfo {author} {\bibfnamefont {A.}~\bibnamefont
  {Shudo}}\ and\ \bibinfo {author} {\bibfnamefont {Y.}~\bibnamefont
  {Shimizu}},\ }\href {\doibase 10.1103/PhysRevA.42.6264} {\bibfield  {journal}
  {\bibinfo  {journal} {Phys. Rev. A}\ }\textbf {\bibinfo {volume} {42}},\
  \bibinfo {pages} {6264} (\bibinfo {year} {1990})}\BibitemShut {NoStop}%
\bibitem [{\citenamefont {Meiss}(1992)}]{Meiss1992}%
  \BibitemOpen
  \bibfield  {author} {\bibinfo {author} {\bibfnamefont {J.~D.}\ \bibnamefont
  {Meiss}},\ }\href {\doibase 10.1063/1.165867} {\bibfield  {journal} {\bibinfo
   {journal} {Chaos}\ }\textbf {\bibinfo {volume} {2}},\ \bibinfo {pages} {267}
  (\bibinfo {year} {1992})}\BibitemShut {NoStop}%
\bibitem [{\citenamefont {Sieber}\ \emph {et~al.}(1993)\citenamefont {Sieber},
  \citenamefont {Smilansky}, \citenamefont {Creagh},\ and\ \citenamefont
  {Littlejohn}}]{Sieber1993}%
  \BibitemOpen
  \bibfield  {author} {\bibinfo {author} {\bibfnamefont {M.}~\bibnamefont
  {Sieber}}, \bibinfo {author} {\bibfnamefont {U.}~\bibnamefont {Smilansky}},
  \bibinfo {author} {\bibfnamefont {S.~C.}\ \bibnamefont {Creagh}}, \ and\
  \bibinfo {author} {\bibfnamefont {R.~G.}\ \bibnamefont {Littlejohn}},\ }\href
  {\doibase 10.1088/0305-4470/26/22/022} {\bibfield  {journal} {\bibinfo
  {journal} {J. Phys. A}\ }\textbf {\bibinfo {volume} {26}},\ \bibinfo {pages}
  {6217} (\bibinfo {year} {1993})}\BibitemShut {NoStop}%
\bibitem [{\citenamefont {Alonso}\ and\ \citenamefont
  {Gaspard}(1994)}]{Alonso1994}%
  \BibitemOpen
  \bibfield  {author} {\bibinfo {author} {\bibfnamefont {D.}~\bibnamefont
  {Alonso}}\ and\ \bibinfo {author} {\bibfnamefont {P.}~\bibnamefont
  {Gaspard}},\ }\href {\doibase doi:10.1088/0305-4470/27/5/023} {\bibfield
  {journal} {\bibinfo  {journal} {J. Phys. A}\ }\textbf {\bibinfo {volume}
  {27}},\ \bibinfo {pages} {1599} (\bibinfo {year} {1994})}\BibitemShut
  {NoStop}%
\bibitem [{\citenamefont {Primack}\ and\ \citenamefont
  {Smilansky}(1994)}]{Primack1994}%
  \BibitemOpen
  \bibfield  {author} {\bibinfo {author} {\bibfnamefont {H.}~\bibnamefont
  {Primack}}\ and\ \bibinfo {author} {\bibfnamefont {U.}~\bibnamefont
  {Smilansky}},\ }\href {\doibase 10.1088/0305-4470/27/13/018} {\bibfield
  {journal} {\bibinfo  {journal} {J. Phys. A}\ }\textbf {\bibinfo {volume}
  {27}},\ \bibinfo {pages} {4439} (\bibinfo {year} {1994})}\BibitemShut
  {NoStop}%
\bibitem [{\citenamefont {Tanner}(1997)}]{Tanner1997}%
  \BibitemOpen
  \bibfield  {author} {\bibinfo {author} {\bibfnamefont {G.}~\bibnamefont
  {Tanner}},\ }\href {\doibase 10.1088/0305-4470/30/8/028} {\bibfield
  {journal} {\bibinfo  {journal} {J. Phys. A}\ }\textbf {\bibinfo {volume}
  {30}},\ \bibinfo {pages} {2863} (\bibinfo {year} {1997})}\BibitemShut
  {NoStop}%
\bibitem [{\citenamefont {Casati}\ and\ \citenamefont
  {Prosen}(1999)}]{Casati1999}%
  \BibitemOpen
  \bibfield  {author} {\bibinfo {author} {\bibfnamefont {G.}~\bibnamefont
  {Casati}}\ and\ \bibinfo {author} {\bibfnamefont {T.}~\bibnamefont
  {Prosen}},\ }\href {\doibase 10.1103/PhysRevE.59.R2516} {\bibfield  {journal}
  {\bibinfo  {journal} {Phys. Rev. E}\ }\textbf {\bibinfo {volume} {59}},\
  \bibinfo {pages} {R2516} (\bibinfo {year} {1999})}\BibitemShut {NoStop}%
\bibitem [{\citenamefont {St{\"o}ckmann}\ and\ \citenamefont
  {Stein}(1990)}]{Stoeckmann1990}%
  \BibitemOpen
  \bibfield  {author} {\bibinfo {author} {\bibfnamefont {H.-J.}\ \bibnamefont
  {St{\"o}ckmann}}\ and\ \bibinfo {author} {\bibfnamefont {J.}~\bibnamefont
  {Stein}},\ }\href {\doibase 10.1103/PhysRevLett.64.2215} {\bibfield
  {journal} {\bibinfo  {journal} {Phys. Rev. Lett.}\ }\textbf {\bibinfo
  {volume} {64}},\ \bibinfo {pages} {2215} (\bibinfo {year}
  {1990})}\BibitemShut {NoStop}%
\bibitem [{\citenamefont {Gr{\"a}f}\ \emph {et~al.}(1992)\citenamefont
  {Gr{\"a}f}, \citenamefont {Harney}, \citenamefont {Lengeler}, \citenamefont
  {Lewenkopf}, \citenamefont {Rangacharyulu}, \citenamefont {Richter},
  \citenamefont {Schardt},\ and\ \citenamefont {Weidenm{\"u}ller}}]{Graf1992}%
  \BibitemOpen
  \bibfield  {author} {\bibinfo {author} {\bibfnamefont {H.-D.}\ \bibnamefont
  {Gr{\"a}f}}, \bibinfo {author} {\bibfnamefont {H.~L.}\ \bibnamefont
  {Harney}}, \bibinfo {author} {\bibfnamefont {H.}~\bibnamefont {Lengeler}},
  \bibinfo {author} {\bibfnamefont {C.~H.}\ \bibnamefont {Lewenkopf}}, \bibinfo
  {author} {\bibfnamefont {C.}~\bibnamefont {Rangacharyulu}}, \bibinfo {author}
  {\bibfnamefont {A.}~\bibnamefont {Richter}}, \bibinfo {author} {\bibfnamefont
  {P.}~\bibnamefont {Schardt}}, \ and\ \bibinfo {author} {\bibfnamefont
  {H.~A.}\ \bibnamefont {Weidenm{\"u}ller}},\ }\href {\doibase
  10.1103/PhysRevLett.69.1296} {\bibfield  {journal} {\bibinfo  {journal}
  {Phys. Rev. Lett.}\ }\textbf {\bibinfo {volume} {69}},\ \bibinfo {pages}
  {1296} (\bibinfo {year} {1992})}\BibitemShut {NoStop}%
\bibitem [{\citenamefont {Marcus}\ \emph {et~al.}(1992)\citenamefont {Marcus},
  \citenamefont {Rimberg}, \citenamefont {Westervelt}, \citenamefont
  {Hopkins},\ and\ \citenamefont {Gossard}}]{Marcus1992}%
  \BibitemOpen
  \bibfield  {author} {\bibinfo {author} {\bibfnamefont {C.~M.}\ \bibnamefont
  {Marcus}}, \bibinfo {author} {\bibfnamefont {A.~J.}\ \bibnamefont {Rimberg}},
  \bibinfo {author} {\bibfnamefont {R.~M.}\ \bibnamefont {Westervelt}},
  \bibinfo {author} {\bibfnamefont {P.~F.}\ \bibnamefont {Hopkins}}, \ and\
  \bibinfo {author} {\bibfnamefont {A.~C.}\ \bibnamefont {Gossard}},\ }\href
  {\doibase 10.1103/PhysRevLett.69.506} {\bibfield  {journal} {\bibinfo
  {journal} {Phys. Rev. Lett.}\ }\textbf {\bibinfo {volume} {69}},\ \bibinfo
  {pages} {506} (\bibinfo {year} {1992})}\BibitemShut {NoStop}%
\bibitem [{\citenamefont {Alt}\ \emph {et~al.}(1995)\citenamefont {Alt},
  \citenamefont {Gr\"af}, \citenamefont {Harney}, \citenamefont {Hofferbert},
  \citenamefont {Lengeler}, \citenamefont {Richter}, \citenamefont {Schardt},\
  and\ \citenamefont {Weidenm\"uller}}]{Alt1995}%
  \BibitemOpen
  \bibfield  {author} {\bibinfo {author} {\bibfnamefont {H.}~\bibnamefont
  {Alt}}, \bibinfo {author} {\bibfnamefont {H.~D.}\ \bibnamefont {Gr\"af}},
  \bibinfo {author} {\bibfnamefont {H.~L.}\ \bibnamefont {Harney}}, \bibinfo
  {author} {\bibfnamefont {R.}~\bibnamefont {Hofferbert}}, \bibinfo {author}
  {\bibfnamefont {H.}~\bibnamefont {Lengeler}}, \bibinfo {author}
  {\bibfnamefont {A.}~\bibnamefont {Richter}}, \bibinfo {author} {\bibfnamefont
  {P.}~\bibnamefont {Schardt}}, \ and\ \bibinfo {author} {\bibfnamefont
  {H.~A.}\ \bibnamefont {Weidenm\"uller}},\ }\href {\doibase
  10.1103/PhysRevLett.74.62} {\bibfield  {journal} {\bibinfo  {journal} {Phys.
  Rev. Lett.}\ }\textbf {\bibinfo {volume} {74}},\ \bibinfo {pages} {62}
  (\bibinfo {year} {1995})}\BibitemShut {NoStop}%
\bibitem [{\citenamefont {Stein}\ \emph {et~al.}(1995)\citenamefont {Stein},
  \citenamefont {St{\"o}ckmann},\ and\ \citenamefont {Stoffregen}}]{Stein1995}%
  \BibitemOpen
  \bibfield  {author} {\bibinfo {author} {\bibfnamefont {J.}~\bibnamefont
  {Stein}}, \bibinfo {author} {\bibfnamefont {H.-J.}\ \bibnamefont
  {St{\"o}ckmann}}, \ and\ \bibinfo {author} {\bibfnamefont {U.}~\bibnamefont
  {Stoffregen}},\ }\href {\doibase 10.1103/PhysRevLett.75.53} {\bibfield
  {journal} {\bibinfo  {journal} {Phys. Rev. Lett.}\ }\textbf {\bibinfo
  {volume} {75}},\ \bibinfo {pages} {53} (\bibinfo {year} {1995})}\BibitemShut
  {NoStop}%
\bibitem [{\citenamefont {Crommie}\ \emph {et~al.}(1995)\citenamefont
  {Crommie}, \citenamefont {Lutz}, \citenamefont {Eigler},\ and\ \citenamefont
  {Heller}}]{Crommie1995}%
  \BibitemOpen
  \bibfield  {author} {\bibinfo {author} {\bibfnamefont {M.~F.}\ \bibnamefont
  {Crommie}}, \bibinfo {author} {\bibfnamefont {C.~P.}\ \bibnamefont {Lutz}},
  \bibinfo {author} {\bibfnamefont {D.~M.}\ \bibnamefont {Eigler}}, \ and\
  \bibinfo {author} {\bibfnamefont {E.~J.}\ \bibnamefont {Heller}},\ }\href
  {\doibase 10.1016/0167-2789(94)00254-N} {\bibfield  {journal} {\bibinfo
  {journal} {Physica D}\ }\textbf {\bibinfo {volume} {83}},\ \bibinfo {pages}
  {98} (\bibinfo {year} {1995})}\BibitemShut {NoStop}%
\bibitem [{\citenamefont {Arcos}\ \emph {et~al.}(1998)\citenamefont {Arcos},
  \citenamefont {B\'aez}, \citenamefont {Cuatl\'ayol}, \citenamefont {Prian},
  \citenamefont {M\'endez-S\'anchez},\ and\ \citenamefont
  {Hern\'andez-Salda\~{n}a}}]{Arcos1998}%
  \BibitemOpen
  \bibfield  {author} {\bibinfo {author} {\bibfnamefont {E.}~\bibnamefont
  {Arcos}}, \bibinfo {author} {\bibfnamefont {G.}~\bibnamefont {B\'aez}},
  \bibinfo {author} {\bibfnamefont {P.~A.}\ \bibnamefont {Cuatl\'ayol}},
  \bibinfo {author} {\bibfnamefont {M.~L.~H.}\ \bibnamefont {Prian}}, \bibinfo
  {author} {\bibfnamefont {R.~A.}\ \bibnamefont {M\'endez-S\'anchez}}, \ and\
  \bibinfo {author} {\bibfnamefont {H.}~\bibnamefont
  {Hern\'andez-Salda\~{n}a}},\ }\href {\doibase 10.1119/1.18913} {\bibfield
  {journal} {\bibinfo  {journal} {Am. J. Phys.}\ }\textbf {\bibinfo {volume}
  {66}},\ \bibinfo {pages} {601} (\bibinfo {year} {1998})}\BibitemShut
  {NoStop}%
\bibitem [{\citenamefont {Friedman}\ \emph {et~al.}(2001)\citenamefont
  {Friedman}, \citenamefont {Kaplan}, \citenamefont {Carasso},\ and\
  \citenamefont {Davidson}}]{Friedman2001}%
  \BibitemOpen
  \bibfield  {author} {\bibinfo {author} {\bibfnamefont {N.}~\bibnamefont
  {Friedman}}, \bibinfo {author} {\bibfnamefont {A.}~\bibnamefont {Kaplan}},
  \bibinfo {author} {\bibfnamefont {D.}~\bibnamefont {Carasso}}, \ and\
  \bibinfo {author} {\bibfnamefont {N.}~\bibnamefont {Davidson}},\ }\href
  {\doibase 10.1103/PhysRevLett.86.1518} {\bibfield  {journal} {\bibinfo
  {journal} {Phys. Rev. Lett.}\ }\textbf {\bibinfo {volume} {86}},\ \bibinfo
  {pages} {1518} (\bibinfo {year} {2001})}\BibitemShut {NoStop}%
\bibitem [{\citenamefont {Fukushima}\ and\ \citenamefont
  {Harayama}(2004)}]{Fukushima2004}%
  \BibitemOpen
  \bibfield  {author} {\bibinfo {author} {\bibfnamefont {T.}~\bibnamefont
  {Fukushima}}\ and\ \bibinfo {author} {\bibfnamefont {T.}~\bibnamefont
  {Harayama}},\ }\href {\doibase 10.1109/JSTQE.2004.836003} {\bibfield
  {journal} {\bibinfo  {journal} {IEEE J. Sel. Top. Quant. Electron.}\ }\textbf
  {\bibinfo {volume} {10}},\ \bibinfo {pages} {1039} (\bibinfo {year}
  {2004})}\BibitemShut {NoStop}%
\bibitem [{\citenamefont {Lebental}\ \emph {et~al.}(2006)\citenamefont
  {Lebental}, \citenamefont {Lauret}, \citenamefont {Hierle},\ and\
  \citenamefont {Zyss}}]{Lebental2006}%
  \BibitemOpen
  \bibfield  {author} {\bibinfo {author} {\bibfnamefont {M.}~\bibnamefont
  {Lebental}}, \bibinfo {author} {\bibfnamefont {J.~S.}\ \bibnamefont
  {Lauret}}, \bibinfo {author} {\bibfnamefont {R.}~\bibnamefont {Hierle}}, \
  and\ \bibinfo {author} {\bibfnamefont {J.}~\bibnamefont {Zyss}},\ }\href
  {\doibase 10.1063/1.2159099} {\bibfield  {journal} {\bibinfo  {journal}
  {Appl. Phys. Lett.}\ }\textbf {\bibinfo {volume} {88}},\ \bibinfo {pages}
  {031108} (\bibinfo {year} {2006})}\BibitemShut {NoStop}%
\bibitem [{\citenamefont {Shinohara}\ \emph {et~al.}(2008)\citenamefont
  {Shinohara}, \citenamefont {Fukushima},\ and\ \citenamefont
  {Harayama}}]{Shinohara2008}%
  \BibitemOpen
  \bibfield  {author} {\bibinfo {author} {\bibfnamefont {S.}~\bibnamefont
  {Shinohara}}, \bibinfo {author} {\bibfnamefont {T.}~\bibnamefont
  {Fukushima}}, \ and\ \bibinfo {author} {\bibfnamefont {T.}~\bibnamefont
  {Harayama}},\ }\href {\doibase 10.1103/PhysRevA.77.033807} {\bibfield
  {journal} {\bibinfo  {journal} {Phys. Rev. A}\ }\textbf {\bibinfo {volume}
  {77}},\ \bibinfo {pages} {033807} (\bibinfo {year} {2008})}\BibitemShut
  {NoStop}%
\bibitem [{\citenamefont {Djellali}\ \emph {et~al.}(2009)\citenamefont
  {Djellali}, \citenamefont {Gozhyk}, \citenamefont {Owens}, \citenamefont
  {Lozenko}, \citenamefont {Lebental}, \citenamefont {Lautru}, \citenamefont
  {Ulysse}, \citenamefont {Kippelen},\ and\ \citenamefont
  {Zyss}}]{Djellali2009}%
  \BibitemOpen
  \bibfield  {author} {\bibinfo {author} {\bibfnamefont {N.}~\bibnamefont
  {Djellali}}, \bibinfo {author} {\bibfnamefont {I.}~\bibnamefont {Gozhyk}},
  \bibinfo {author} {\bibfnamefont {D.}~\bibnamefont {Owens}}, \bibinfo
  {author} {\bibfnamefont {S.}~\bibnamefont {Lozenko}}, \bibinfo {author}
  {\bibfnamefont {M.}~\bibnamefont {Lebental}}, \bibinfo {author}
  {\bibfnamefont {J.}~\bibnamefont {Lautru}}, \bibinfo {author} {\bibfnamefont
  {C.}~\bibnamefont {Ulysse}}, \bibinfo {author} {\bibfnamefont
  {B.}~\bibnamefont {Kippelen}}, \ and\ \bibinfo {author} {\bibfnamefont
  {J.}~\bibnamefont {Zyss}},\ }\href {\doibase 10.1063/1.3205474} {\bibfield
  {journal} {\bibinfo  {journal} {Appl. Phys. Lett.}\ }\textbf {\bibinfo
  {volume} {95}},\ \bibinfo {pages} {101108} (\bibinfo {year}
  {2009})}\BibitemShut {NoStop}%
\bibitem [{\citenamefont {Jackson}(1999)}]{Jackson1999}%
  \BibitemOpen
  \bibfield  {author} {\bibinfo {author} {\bibfnamefont {J.~D.}\ \bibnamefont
  {Jackson}},\ }\href@noop {} {\emph {\bibinfo {title} {Classical
  Electrodynamics}}}\ (\bibinfo  {publisher} {John Wiley \& Sons, Inc.},\
  \bibinfo {address} {New York},\ \bibinfo {year} {1999})\BibitemShut {NoStop}%
\bibitem [{\citenamefont {Sridhar}\ \emph {et~al.}(1992)\citenamefont
  {Sridhar}, \citenamefont {Hogenboom},\ and\ \citenamefont
  {Willemsen}}]{Sridhar1992}%
  \BibitemOpen
  \bibfield  {author} {\bibinfo {author} {\bibfnamefont {S.}~\bibnamefont
  {Sridhar}}, \bibinfo {author} {\bibfnamefont {D.}~\bibnamefont {Hogenboom}},
  \ and\ \bibinfo {author} {\bibfnamefont {B.}~\bibnamefont {Willemsen}},\
  }\href {\doibase 10.1007/BF01048844} {\bibfield  {journal} {\bibinfo
  {journal} {J. Stat. Phys.}\ }\textbf {\bibinfo {volume} {68}},\ \bibinfo
  {pages} {239} (\bibinfo {year} {1992})}\BibitemShut {NoStop}%
\bibitem [{\citenamefont {Heller}(1984)}]{Heller1984}%
  \BibitemOpen
  \bibfield  {author} {\bibinfo {author} {\bibfnamefont {E.~J.}\ \bibnamefont
  {Heller}},\ }\href {\doibase 10.1103/PhysRevLett.53.1515} {\bibfield
  {journal} {\bibinfo  {journal} {Phys. Rev. Lett.}\ }\textbf {\bibinfo
  {volume} {53}},\ \bibinfo {pages} {1515} (\bibinfo {year}
  {1984})}\BibitemShut {NoStop}%
\bibitem [{\citenamefont {Chern}\ \emph {et~al.}(2004)\citenamefont {Chern},
  \citenamefont {Poon}, \citenamefont {Chang}, \citenamefont {Ben-Messaoud},
  \citenamefont {Alloschery}, \citenamefont {Toussaere}, \citenamefont {Zyss},\
  and\ \citenamefont {Kuo}}]{Chern2004}%
  \BibitemOpen
  \bibfield  {author} {\bibinfo {author} {\bibfnamefont {G.~D.}\ \bibnamefont
  {Chern}}, \bibinfo {author} {\bibfnamefont {A.~W.}\ \bibnamefont {Poon}},
  \bibinfo {author} {\bibfnamefont {R.~K.}\ \bibnamefont {Chang}}, \bibinfo
  {author} {\bibfnamefont {T.}~\bibnamefont {Ben-Messaoud}}, \bibinfo {author}
  {\bibfnamefont {O.}~\bibnamefont {Alloschery}}, \bibinfo {author}
  {\bibfnamefont {E.}~\bibnamefont {Toussaere}}, \bibinfo {author}
  {\bibfnamefont {J.}~\bibnamefont {Zyss}}, \ and\ \bibinfo {author}
  {\bibfnamefont {S.-Y.}\ \bibnamefont {Kuo}},\ }\href {\doibase
  10.1364/OL.29.001674} {\bibfield  {journal} {\bibinfo  {journal} {Opt.
  Lett.}\ }\textbf {\bibinfo {volume} {29}},\ \bibinfo {pages} {1674} (\bibinfo
  {year} {2004})}\BibitemShut {NoStop}%
\bibitem [{\citenamefont {Kim}\ \emph {et~al.}(2011{\natexlab{a}})\citenamefont
  {Kim}, \citenamefont {Park}, \citenamefont {Yi},\ and\ \citenamefont
  {Kim}}]{Kim2011}%
  \BibitemOpen
  \bibfield  {author} {\bibinfo {author} {\bibfnamefont {M.-W.}\ \bibnamefont
  {Kim}}, \bibinfo {author} {\bibfnamefont {K.-W.}\ \bibnamefont {Park}},
  \bibinfo {author} {\bibfnamefont {C.-H.}\ \bibnamefont {Yi}}, \ and\ \bibinfo
  {author} {\bibfnamefont {C.-M.}\ \bibnamefont {Kim}},\ }\href {\doibase
  10.1063/1.3598406} {\bibfield  {journal} {\bibinfo  {journal} {Appl. Phys.
  Lett.}\ }\textbf {\bibinfo {volume} {98}},\ \bibinfo {pages} {241110}
  (\bibinfo {year} {2011}{\natexlab{a}})}\BibitemShut {NoStop}%
\bibitem [{\citenamefont {Kim}\ \emph {et~al.}(2011{\natexlab{b}})\citenamefont
  {Kim}, \citenamefont {Park}, \citenamefont {Yi},\ and\ \citenamefont
  {Kim}}]{Kim2011a}%
  \BibitemOpen
  \bibfield  {author} {\bibinfo {author} {\bibfnamefont {M.-W.}\ \bibnamefont
  {Kim}}, \bibinfo {author} {\bibfnamefont {K.-W.}\ \bibnamefont {Park}},
  \bibinfo {author} {\bibfnamefont {C.-H.}\ \bibnamefont {Yi}}, \ and\ \bibinfo
  {author} {\bibfnamefont {C.-M.}\ \bibnamefont {Kim}},\ }\href {\doibase
  10.1364/OL.36.004503} {\bibfield  {journal} {\bibinfo  {journal} {Opt.
  Lett.}\ }\textbf {\bibinfo {volume} {36}},\ \bibinfo {pages} {4503} (\bibinfo
  {year} {2011}{\natexlab{b}})}\BibitemShut {NoStop}%
\bibitem [{\citenamefont {Shinohara}\ \emph {et~al.}(2011)\citenamefont
  {Shinohara}, \citenamefont {Harayama}, \citenamefont {Fukushima},
  \citenamefont {Hentschel}, \citenamefont {Sunada},\ and\ \citenamefont
  {Narimanov}}]{Shinohara2011b}%
  \BibitemOpen
  \bibfield  {author} {\bibinfo {author} {\bibfnamefont {S.}~\bibnamefont
  {Shinohara}}, \bibinfo {author} {\bibfnamefont {T.}~\bibnamefont {Harayama}},
  \bibinfo {author} {\bibfnamefont {T.}~\bibnamefont {Fukushima}}, \bibinfo
  {author} {\bibfnamefont {M.}~\bibnamefont {Hentschel}}, \bibinfo {author}
  {\bibfnamefont {S.}~\bibnamefont {Sunada}}, \ and\ \bibinfo {author}
  {\bibfnamefont {E.~E.}\ \bibnamefont {Narimanov}},\ }\href {\doibase
  10.1103/PhysRevA.83.053837} {\bibfield  {journal} {\bibinfo  {journal} {Phys.
  Rev. A}\ }\textbf {\bibinfo {volume} {83}},\ \bibinfo {pages} {053837}
  (\bibinfo {year} {2011})}\BibitemShut {NoStop}%
\bibitem [{\citenamefont {Yi}\ \emph {et~al.}(2011)\citenamefont {Yi},
  \citenamefont {Lee}, \citenamefont {Kim}, \citenamefont {Cho}, \citenamefont
  {Lee}, \citenamefont {Lee}, \citenamefont {Wiersig},\ and\ \citenamefont
  {Kim}}]{Yi2011}%
  \BibitemOpen
  \bibfield  {author} {\bibinfo {author} {\bibfnamefont {C.-H.}\ \bibnamefont
  {Yi}}, \bibinfo {author} {\bibfnamefont {S.~H.}\ \bibnamefont {Lee}},
  \bibinfo {author} {\bibfnamefont {M.-W.}\ \bibnamefont {Kim}}, \bibinfo
  {author} {\bibfnamefont {J.}~\bibnamefont {Cho}}, \bibinfo {author}
  {\bibfnamefont {J.}~\bibnamefont {Lee}}, \bibinfo {author} {\bibfnamefont
  {S.-Y.}\ \bibnamefont {Lee}}, \bibinfo {author} {\bibfnamefont
  {J.}~\bibnamefont {Wiersig}}, \ and\ \bibinfo {author} {\bibfnamefont
  {C.-M.}\ \bibnamefont {Kim}},\ }\href {\doibase 10.1103/PhysRevA.84.041803}
  {\bibfield  {journal} {\bibinfo  {journal} {Phys. Rev. A}\ }\textbf {\bibinfo
  {volume} {84}},\ \bibinfo {pages} {041803} (\bibinfo {year}
  {2011})}\BibitemShut {NoStop}%
\bibitem [{\citenamefont {Bittner}\ \emph {et~al.}(2009)\citenamefont
  {Bittner}, \citenamefont {Dietz}, \citenamefont {Miski-Oglu}, \citenamefont
  {{Oria Iriarte}}, \citenamefont {Richter},\ and\ \citenamefont
  {Sch\"{a}fer}}]{Bittner2009}%
  \BibitemOpen
  \bibfield  {author} {\bibinfo {author} {\bibfnamefont {S.}~\bibnamefont
  {Bittner}}, \bibinfo {author} {\bibfnamefont {B.}~\bibnamefont {Dietz}},
  \bibinfo {author} {\bibfnamefont {M.}~\bibnamefont {Miski-Oglu}}, \bibinfo
  {author} {\bibfnamefont {P.}~\bibnamefont {{Oria Iriarte}}}, \bibinfo
  {author} {\bibfnamefont {A.}~\bibnamefont {Richter}}, \ and\ \bibinfo
  {author} {\bibfnamefont {F.}~\bibnamefont {Sch\"{a}fer}},\ }\href {\doibase
  10.1103/PhysRevA.80.023825} {\bibfield  {journal} {\bibinfo  {journal} {Phys.
  Rev. A}\ }\textbf {\bibinfo {volume} {80}},\ \bibinfo {pages} {023825}
  (\bibinfo {year} {2009})}\BibitemShut {NoStop}%
\bibitem [{\citenamefont {Hecht}(2002)}]{Hecht2002}%
  \BibitemOpen
  \bibfield  {author} {\bibinfo {author} {\bibfnamefont {E.}~\bibnamefont
  {Hecht}},\ }\href@noop {} {\emph {\bibinfo {title} {Optics}}},\ \bibinfo
  {edition} {4th}\ ed.\ (\bibinfo  {publisher} {Addison-Wesley},\ \bibinfo
  {address} {San Francisco, CA},\ \bibinfo {year} {2002})\BibitemShut {NoStop}%
\bibitem [{\citenamefont {Biham}\ and\ \citenamefont
  {Kvale}(1992)}]{Biham1992}%
  \BibitemOpen
  \bibfield  {author} {\bibinfo {author} {\bibfnamefont {O.}~\bibnamefont
  {Biham}}\ and\ \bibinfo {author} {\bibfnamefont {M.}~\bibnamefont {Kvale}},\
  }\href {\doibase 10.1103/PhysRevA.46.6334} {\bibfield  {journal} {\bibinfo
  {journal} {Phys. Rev. A}\ }\textbf {\bibinfo {volume} {46}},\ \bibinfo
  {pages} {6334} (\bibinfo {year} {1992})}\BibitemShut {NoStop}%
\bibitem [{\citenamefont {Sieber}\ and\ \citenamefont
  {Steiner}(1990)}]{Sieber1990a}%
  \BibitemOpen
  \bibfield  {author} {\bibinfo {author} {\bibfnamefont {M.}~\bibnamefont
  {Sieber}}\ and\ \bibinfo {author} {\bibfnamefont {F.}~\bibnamefont
  {Steiner}},\ }\href {\doibase 10.1016/0167-2789(90)90058-W} {\bibfield
  {journal} {\bibinfo  {journal} {Physica D}\ }\textbf {\bibinfo {volume}
  {44}},\ \bibinfo {pages} {248} (\bibinfo {year} {1990})}\BibitemShut
  {NoStop}%
\bibitem [{\citenamefont {Lin}\ and\ \citenamefont {Jensen}(1997)}]{Lin1997}%
  \BibitemOpen
  \bibfield  {author} {\bibinfo {author} {\bibfnamefont {W.~A.}\ \bibnamefont
  {Lin}}\ and\ \bibinfo {author} {\bibfnamefont {R.~V.}\ \bibnamefont
  {Jensen}},\ }\href {\doibase 10.1103/PhysRevE.56.5251} {\bibfield  {journal}
  {\bibinfo  {journal} {Phys. Rev. E}\ }\textbf {\bibinfo {volume} {56}},\
  \bibinfo {pages} {5251} (\bibinfo {year} {1997})}\BibitemShut {NoStop}%
\bibitem [{\citenamefont {Robbins}(1989)}]{Robbins1989}%
  \BibitemOpen
  \bibfield  {author} {\bibinfo {author} {\bibfnamefont {J.~M.}\ \bibnamefont
  {Robbins}},\ }\href {\doibase 10.1103/PhysRevA.40.2128} {\bibfield  {journal}
  {\bibinfo  {journal} {Phys. Rev. A}\ }\textbf {\bibinfo {volume} {40}},\
  \bibinfo {pages} {2128} (\bibinfo {year} {1989})}\BibitemShut {NoStop}%
\bibitem [{\citenamefont {Takami}(1995)}]{Takami1995}%
  \BibitemOpen
  \bibfield  {author} {\bibinfo {author} {\bibfnamefont {T.}~\bibnamefont
  {Takami}},\ }\href {\doibase 10.1103/PhysRevE.52.2434} {\bibfield  {journal}
  {\bibinfo  {journal} {Phys. Rev. E}\ }\textbf {\bibinfo {volume} {52}},\
  \bibinfo {pages} {2434} (\bibinfo {year} {1995})}\BibitemShut {NoStop}%
\bibitem [{\citenamefont {Bruus}\ and\ \citenamefont
  {Whelan}(1996)}]{Bruus1996}%
  \BibitemOpen
  \bibfield  {author} {\bibinfo {author} {\bibfnamefont {H.}~\bibnamefont
  {Bruus}}\ and\ \bibinfo {author} {\bibfnamefont {N.~D.}\ \bibnamefont
  {Whelan}},\ }\href {\doibase 10.1088/0951-7715/9/4/012} {\bibfield  {journal}
  {\bibinfo  {journal} {Nonlinearity}\ }\textbf {\bibinfo {volume} {9}},\
  \bibinfo {pages} {1023} (\bibinfo {year} {1996})}\BibitemShut {NoStop}%
\bibitem [{\citenamefont {Berry}\ and\ \citenamefont
  {Mount}(1972)}]{Berry1972}%
  \BibitemOpen
  \bibfield  {author} {\bibinfo {author} {\bibfnamefont {M.~V.}\ \bibnamefont
  {Berry}}\ and\ \bibinfo {author} {\bibfnamefont {K.~E.}\ \bibnamefont
  {Mount}},\ }\href {\doibase 10.1088/0034-4885/35/1/306} {\bibfield  {journal}
  {\bibinfo  {journal} {Rep. Prog. Phys.}\ }\textbf {\bibinfo {volume} {35}},\
  \bibinfo {pages} {315} (\bibinfo {year} {1972})}\BibitemShut {NoStop}%
\bibitem [{\citenamefont {D\'{e}canini}\ and\ \citenamefont
  {Folacci}(2003)}]{Decanini2003}%
  \BibitemOpen
  \bibfield  {author} {\bibinfo {author} {\bibfnamefont {Y.}~\bibnamefont
  {D\'{e}canini}}\ and\ \bibinfo {author} {\bibfnamefont {A.}~\bibnamefont
  {Folacci}},\ }\href {\doibase 10.1103/PhysRevE.68.046204} {\bibfield
  {journal} {\bibinfo  {journal} {Phys. Rev. E}\ }\textbf {\bibinfo {volume}
  {68}},\ \bibinfo {pages} {046204} (\bibinfo {year} {2003})}\BibitemShut
  {NoStop}%
\bibitem [{\citenamefont {Abramowitz}\ and\ \citenamefont
  {Stegun}(1972)}]{Abramowitz1972}%
  \BibitemOpen
  \bibfield  {author} {\bibinfo {author} {\bibfnamefont {M.}~\bibnamefont
  {Abramowitz}}\ and\ \bibinfo {author} {\bibfnamefont {I.~A.}\ \bibnamefont
  {Stegun}},\ }\href@noop {} {\emph {\bibinfo {title} {Handbook of Mathematical
  Functions}}}\ (\bibinfo  {publisher} {Dover Publications},\ \bibinfo
  {address} {New York},\ \bibinfo {year} {1972})\BibitemShut {NoStop}%
\bibitem [{\citenamefont {Hentschel}\ and\ \citenamefont
  {Schomerus}(2002)}]{Hentschel2002}%
  \BibitemOpen
  \bibfield  {author} {\bibinfo {author} {\bibfnamefont {M.}~\bibnamefont
  {Hentschel}}\ and\ \bibinfo {author} {\bibfnamefont {H.}~\bibnamefont
  {Schomerus}},\ }\href {\doibase 10.1103/PhysRevE.65.045603} {\bibfield
  {journal} {\bibinfo  {journal} {Phys. Rev. E}\ }\textbf {\bibinfo {volume}
  {65}},\ \bibinfo {pages} {045603} (\bibinfo {year} {2002})}\BibitemShut
  {NoStop}%
\bibitem [{\citenamefont {Wiersig}(2002)}]{Wiersig2002a}%
  \BibitemOpen
  \bibfield  {author} {\bibinfo {author} {\bibfnamefont {J.}~\bibnamefont
  {Wiersig}},\ }\href {\doibase 10.1088/1464-4258/5/1/308} {\bibfield
  {journal} {\bibinfo  {journal} {J. Opt. A}\ }\textbf {\bibinfo {volume}
  {5}},\ \bibinfo {pages} {53} (\bibinfo {year} {2002})}\BibitemShut {NoStop}%
\bibitem [{\citenamefont {Schomerus}\ and\ \citenamefont
  {Sieber}(1997)}]{Schomerus1997}%
  \BibitemOpen
  \bibfield  {author} {\bibinfo {author} {\bibfnamefont {H.}~\bibnamefont
  {Schomerus}}\ and\ \bibinfo {author} {\bibfnamefont {M.}~\bibnamefont
  {Sieber}},\ }\href {\doibase 10.1088/0305-4470/30/13/010} {\bibfield
  {journal} {\bibinfo  {journal} {J. Phys. A}\ }\textbf {\bibinfo {volume}
  {30}},\ \bibinfo {pages} {4537} (\bibinfo {year} {1997})}\BibitemShut
  {NoStop}%
\bibitem [{\citenamefont {Lee}\ \emph {et~al.}(2009)\citenamefont {Lee},
  \citenamefont {Yang}, \citenamefont {Moon}, \citenamefont {Lee},
  \citenamefont {Shim}, \citenamefont {Kim}, \citenamefont {Lee},\ and\
  \citenamefont {An}}]{Lee2009}%
  \BibitemOpen
  \bibfield  {author} {\bibinfo {author} {\bibfnamefont {S.-B.}\ \bibnamefont
  {Lee}}, \bibinfo {author} {\bibfnamefont {J.}~\bibnamefont {Yang}}, \bibinfo
  {author} {\bibfnamefont {S.}~\bibnamefont {Moon}}, \bibinfo {author}
  {\bibfnamefont {S.-Y.}\ \bibnamefont {Lee}}, \bibinfo {author} {\bibfnamefont
  {J.-B.}\ \bibnamefont {Shim}}, \bibinfo {author} {\bibfnamefont {S.~W.}\
  \bibnamefont {Kim}}, \bibinfo {author} {\bibfnamefont {J.-H.}\ \bibnamefont
  {Lee}}, \ and\ \bibinfo {author} {\bibfnamefont {K.}~\bibnamefont {An}},\
  }\href {\doibase 10.1103/PhysRevA.80.011802} {\bibfield  {journal} {\bibinfo
  {journal} {Phys. Rev. A}\ }\textbf {\bibinfo {volume} {80}},\ \bibinfo
  {pages} {011802} (\bibinfo {year} {2009})}\BibitemShut {NoStop}%
\bibitem [{\citenamefont {Altmann}\ \emph {et~al.}(2008)\citenamefont
  {Altmann}, \citenamefont {Del~Magno},\ and\ \citenamefont
  {Hentschel}}]{Altmann2008c}%
  \BibitemOpen
  \bibfield  {author} {\bibinfo {author} {\bibfnamefont {E.~G.}\ \bibnamefont
  {Altmann}}, \bibinfo {author} {\bibfnamefont {G.}~\bibnamefont {Del~Magno}},
  \ and\ \bibinfo {author} {\bibfnamefont {M.}~\bibnamefont {Hentschel}},\
  }\href {\doibase 10.1209/0295-5075/84/10008} {\bibfield  {journal} {\bibinfo
  {journal} {Europhys. Lett.}\ }\textbf {\bibinfo {volume} {84}},\ \bibinfo
  {pages} {10008} (\bibinfo {year} {2008})}\BibitemShut {NoStop}%
\bibitem [{\citenamefont {Unterhinninghofen}\ and\ \citenamefont
  {Wiersig}(2010)}]{Unterhinninghofen2010a}%
  \BibitemOpen
  \bibfield  {author} {\bibinfo {author} {\bibfnamefont {J.}~\bibnamefont
  {Unterhinninghofen}}\ and\ \bibinfo {author} {\bibfnamefont {J.}~\bibnamefont
  {Wiersig}},\ }\href {\doibase 10.1103/PhysRevE.82.026202} {\bibfield
  {journal} {\bibinfo  {journal} {Phys. Rev. E}\ }\textbf {\bibinfo {volume}
  {82}},\ \bibinfo {pages} {026202} (\bibinfo {year} {2010})}\BibitemShut
  {NoStop}%
\end{thebibliography}
\end{document}